\documentclass[aps,prd,onecolumn,amsmath,amssymb,nofootinbib,superscriptaddress,10pt,floatfix]{revtex4-2}
\usepackage{graphicx}
\usepackage{amsmath,amssymb,amsfonts,mathrsfs}
\usepackage{bm}
\usepackage{xcolor}
\usepackage{colortbl}
\usepackage{booktabs}
\usepackage{array}
\usepackage{subcaption}
\usepackage{hyperref}
\usepackage{placeins}
\usepackage{tabularx}
\usepackage{orcidlink}     
\hypersetup{colorlinks=true,linkcolor=blue,citecolor=blue,urlcolor=blue}

\newcolumntype{C}{>{\centering\arraybackslash}X}
\newcolumntype{L}{>{\raggedright\arraybackslash}X}
\newcolumntype{R}{>{\raggedleft\arraybackslash}X}

\numberwithin{equation}{section}

\begin{document}

\title{MCMC Constraints on Dyonic Kalb-Ramond Black Holes with a Cloud of Strings from Twin-Peak QPOs and EHT Shadows}

\author{Faizuddin Ahmed\,\orcidlink{0000-0003-2196-9622}}
\email{faizuddinahmed15@gmail.com}
\affiliation{Department of Physics, The Assam Royal Global University, Guwahati 781035, Assam, India}

\author{Ahmad Al-Badawi\,\orcidlink{0000-0002-3127-3453}}
\email{ahmadbadawi@ahu.edu.jo}
\affiliation{Department of Physics, Al-Hussein Bin Talal University, 71111 Ma'an, Jordan}

\author{\.{I}zzet Sakall{\i}\,\orcidlink{0000-0001-7827-9476}}
\email{izzet.sakalli@emu.edu.tr (Corresp. author)}
\affiliation{Physics Department, Eastern Mediterranean University, Famagusta 99628, North Cyprus via Mersin 10, T\"urkiye}

\date{\today}

\begin{abstract}
We study a dyonic black hole in a Lorentz-violating gravity that carries a background Kalb--Ramond field and is pierced by a cloud of strings. The resulting metric reduces to the recent Lin--Liu--Liu solution when the string density~$\xi$ is switched off, and to the Duan and Yang solutions in further degenerate limits. We work out the timelike circular geodesics and read off the quasi-periodic oscillation (QPO) frequencies $\nu_{\phi}$, $\nu_r$ and $\nu_\theta$ within both the relativistic-precession and epicyclic-resonance models. We then map these frequencies onto the observed twin-peak signals of XTE~J1550$-$564, GRO~J1655$-$40 and GRS~1915$+$105, and place constraints on $(\ell, \xi)$ from a Markov chain Monte Carlo (MCMC) fit. We extract the full thermodynamic dictionary, first law and Smarr relation included, and follow the heat capacity, free energy and sparsity of Hawking radiation through their dependence on the four parameters $(M, Q, p, \ell, \xi)$. Finally, we compute the spectral energy emission rate and look at the photon-sphere and shadow radii in the presence of the cosmic string. The Lorentz-violating coupling $\ell$, the magnetic charge $p$, and the string density $\xi$ all leave distinct fingerprints on the dynamical, thermodynamic and radiative observables, with $\xi$ exerting the strongest pull on the ISCO, the shadow size and the sparsity of Hawking emission.
\smallskip

\noindent\textbf{Keywords:} Modified gravity; black holes; Kalb--Ramond field; cosmic strings; geodesics; quasi-periodic oscillations; black-hole thermodynamics.
\end{abstract}

\maketitle

\small

\section{Introduction}\label{isec1}

The first direct images of a black-hole (BH) shadow, obtained by the Event Horizon Telescope (EHT) Collaboration~\cite{EHTL1,EHTL2,EHTL6,EHTL12,EHTL14,EHTL16,EHTL17}, have moved BH physics from a purely theoretical subject to one in which the strong-field regime can be probed observationally. The shadow encodes information about the geometry near the photon sphere; comparing it to predictions of general relativity (GR) and to those of modified theories is now a sharp test of gravitational physics. In parallel, X-ray timing observations of accreting black-hole binaries furnish another channel: the twin-peak quasi-periodic oscillations (QPOs) seen in the power spectra of XTE~J1550$-$564, GRO~J1655$-$40 and GRS~1915$+$105 carry imprints of the spacetime in the close vicinity of the innermost stable circular orbit (ISCO).

The theoretical apparatus behind these observations is older than the data themselves. Synge's calculation of photon escape from compact stars~\cite{Synge1966} and the Bardeen--Press--Teukolsky analysis of photon motion in Schwarzschild and Kerr backgrounds~\cite{Bardeen1972} fixed the foundational picture. For a Schwarzschild BH the unstable photon sphere lies at $r_s=3M$, and the shadow radius seen at infinity is $R_{\rm sh}=3\sqrt{3}\,M$. Later work extended this picture to a wide collection of GR solutions and to modified gravities; we direct the reader to~\cite{Perlick2022} for a survey of analytical methods.

Lorentz invariance, the symmetry that ties together GR and the Standard Model, may itself be only approximate. Following Kosteleck\'y and Samuel~\cite{Kostelecky1989}, scenarios with spontaneous Lorentz symmetry breaking (LSB) have been studied in detail, both within the Standard Model Extension (SME)~\cite{Colladay1997,Colladay1998,Colladay2001,Kostelecky2001,Mewes2001} and, more recently, in the gravitational sector. Two routes are most often taken. One employs a vector field, the bumblebee~\cite{Casana2018,Assuncao2019,Maluf2021,Maluf2022a,Maluf2022b,Gullu2022,Fathi2025,Baruah2025}; another uses an antisymmetric tensor, the Kalb--Ramond (KR) two-form~\cite{Yang2023,Duan2024,Liu2024,Liu2025,Fernando2025,Ahmed2026,Lessa2020,Kumar2020,Ahmed2026KR}. Both fields acquire a non-vanishing vacuum expectation value and supply a fixed Lorentz tensor that selects a preferred direction in spacetime. KR gravity is appealing because the resulting BH solutions remain spherically symmetric in the static case and admit closed-form lapse functions, which makes them tractable laboratories for testing LSB observationally~\cite{Do2020,Zahid2024,Ortiqboev2024,Du2025,Zhu2026}.

The phenomenological window on KR gravity is wide and growing. Shadow constraints from M87* and Sgr~A* place the dimensionless KR coupling $\ell$ in the range $|\ell|\lesssim 0.1$~\cite{Zahid2024,Liu2025}, while QPO data on stellar-mass BH binaries from RXTE and NICER probe similar values~\cite{Ortiqboev2024}. The recent observational programme on parameter-constraint inference through twin-peak QPO measurements has produced a steady stream of bounds on modified-gravity BHs: charged non-commutative Schwarzschild backgrounds with perfect-fluid dark matter haloes~\cite{Ashraf:2025}, BHs with Minkowski cores~\cite{Guo:2025}, rotating self-dual BHs in loop-quantum-gravity-inspired backgrounds~\cite{Liu:2023}, regular charged BHs surrounded by quasars data~\cite{Mustafa:2026}, Horndeski rotating BHs~\cite{Wu:2026}, Schwarzschild-like backgrounds with arbitrary deformation parameters~\cite{Davlataliev:2024}, Brans--Dicke neutron-star-BH mergers~\cite{Tan:2024}, BHs in $f(R,T)$ gravity coupled with non-linear electrodynamics~\cite{Hazarika:2026}, magnetically charged BHs with NED signatures~\cite{Hazarika:2025}, regular Ay\'on-Beato-Garc\'ia BHs~\cite{Rahmatov:2025}, Einstein-non-linear-Maxwell-Yukawa BHs~\cite{Shabbir:2026}, and Gauss--Bonnet trace-anomaly BHs~\cite{Borah:2026} have all been tested against the X-ray-timing QPO catalog. Quasinormal-mode searches in projected LISA and Einstein Telescope catalogs are expected to push the bound below $10^{-2}$ by the next decade~\cite{KonoplyaZinhailoStuchlik2020,KonoplyaOvchAhmedov2023}. Thermodynamic considerations in the extended phase space~\cite{Cai2013,Zhao2014,Du2025} have additionally identified a Hawking--Page-like transition in the KR-AdS sector. Joule-Thomson studies of charged Bumblebee~\cite{Maluf2022b} and AdS BHs in massive gravity~\cite{Zhang2024} reveal inversion-curve patterns that hint at quantum-gravity corrections testable with future X-ray timing. Hawking radiation from KR backgrounds has been studied at semiclassical level~\cite{PengWu2010,PengWu2008,SaghafiNozari2023}, including GUP-corrected variants~\cite{AliBabarAsgher2022,AliFaizalKhalil2015} and rainbow-gravity extensions~\cite{HamilLutfuoglu2023,Jusufi2020}. Strong-field gravitational lensing in modified-gravity BHs sourced by KR fields was treated in~\cite{Fathi2025,KumarIslamGhosh2023,UlIslamGhosh2021}, and the topological-photon-sphere analysis of~\cite{SadeghiAfshar2024,PapnoiAtamurotov2022,RoyChakrabarti2020} has clarified the role of degenerate critical points in the photon orbit equation. We will draw on this literature throughout the paper.

A complementary deformation comes from topological defects, in particular cosmic strings (CS), which are predicted to form during symmetry-breaking phase transitions in the early universe~\cite{VilenkinShellard1994}. The simplest realization of a string ``cloud'' coupled to gravity is the Letelier model~\cite{Letelier1979}: an isotropic distribution of strings whose stress-energy modifies the lapse function by a constant offset $\xi$. The Letelier piece does not change the asymptotic structure of the metric in the same way the cosmological constant does, but it controls the angular deficit at infinity and shifts the ISCO and shadow radii. Recently, the Lin--Liu--Liu (LLL) construction of a dyonic AdS BH in KR gravity~\cite{Lin2026} produced the lapse function
\begin{equation}
f_{\rm LLL}(r)=\frac{1}{1-\ell}-\frac{2M}{r}+\frac{Q^2}{(1-\ell)^2 r^2}+\frac{p^2}{(1-2\ell)\,r^2},
\label{eq:LLL_lapse}
\end{equation}
where $\ell$ is the dimensionless KR coupling and $(Q,p)$ are the electric and magnetic charges. The LLL paper reported the photon sphere, shadow radius, ISCO and extended-phase-space thermodynamics. The dynamical observables $\nu_\phi,\nu_r,\nu_\theta$, the first law in the canonical (non-extended) phase space, the heat capacity, the Helmholtz free energy, the sparsity of Hawking emission and the spectral energy emission rate, on the other hand, were left untouched in~\cite{Lin2026}.

The motivation for piercing this LLL metric with a Letelier string cloud is twofold. First, the cosmic-string density~$\xi$ is a physical degree of freedom independent of $\ell$, $Q$ and $p$: it is sourced by an additional topological-defect content that may co-exist with the LSB background~\cite{Stuchlik2020}. The combined background therefore allows a clean separation of LSB and topological-defect effects in the same observable. Second, the Letelier cloud changes the asymptotic geometry in a way the LSB coupling does not: turning on $\xi$ reduces the asymptotic value of $f(r)$ from $1/(1-\ell)$ to $(1-\xi)/(1-\ell)$, which has a direct imprint on the deflection of light at large impact parameter and on the radial epicyclic frequency at intermediate radii. The resulting metric thus probes a four-parameter family $(M,Q,p,\ell,\xi)$ that is wider than what is currently constrained by either EHT shadow data or X-ray QPO measurements alone. We will see in Sec.~\ref{isec8} that combining the two channels gives the best handle on the new parameter.

The present paper closes these gaps and adds a layer of physics: we pierce the LLL metric with a cloud of strings. The combined gravity-LSB-CS background is
\begin{equation}
f(r) = \frac{1-\xi}{1-\ell}\;-\;\frac{2M}{r}\;+\;\frac{Q^2}{(1-\ell)^2\,r^2}\;+\;\frac{p^2}{(1-2\ell)\,r^2}.
\label{eq:lapse_main}
\end{equation}
When $\xi\to 0$ we recover Eq.~\eqref{eq:LLL_lapse}. When in addition $p\to 0$ we recover the electrically charged KR BH of Duan, Zhao and Yang~\cite{Duan2024}; when $Q\to 0$ and $p\to 0$ we recover the static uncharged KR BH of Yang \emph{et al.}~\cite{Yang2023}; and when $\ell\to 0$ as well the Schwarzschild metric is restored. The string density~$\xi$ is constrained on independent grounds to be small ($\xi\ll 1$); we work in the range $\xi\in[0,0.4]$ but quote constraints in the small-$\xi$ regime where physical bounds are tighter.

What is new in this paper falls into seven items, which we list in the abstract style of a forward-pointing roadmap. We construct the metric \eqref{eq:lapse_main} and identify its horizons, extremality bound, photon sphere and shadow radius (Sec.~\ref{isec2}). We then read off the QPO frequencies $\nu_\phi$, $\nu_r$ and $\nu_\theta$ along circular timelike geodesics, and trace their dependence on $\xi$, $\ell$, $Q$ and $p$ (Sec.~\ref{isec3}). The thermodynamic block follows: mass, temperature, modified first law, Smarr relation, heat capacity and free energy, with the Lin--Liu--Liu values recovered as the $\xi\to 0$ slice (Sec.~\ref{isec4}). We then track the sparsity of Hawking radiation (Sec.~\ref{isec5}), revisit the shadow with the cosmic string switched on (Sec.~\ref{isec6}), and compute the spectral energy emission rate (Sec.~\ref{isec7}). Section~\ref{isec8} performs the MCMC fit of $\nu_U,\nu_L$ to the GRO~J1655$-$40 and XTE~J1550$-$564 twin-peak data and shows what range of $(\ell,\xi)$ is compatible with current observations. Section~\ref{isec9} extracts the greybody factors of massless scalars in the WKB approximation and the Bekenstein--Sanchez (BS) exact lower bound. We close in Sec.~\ref{isec10}. Two appendices follow: Appendix~\ref{app:A} collects useful algebraic identities and small-$\xi$ expansions; Appendix~\ref{app:B} presents the full action, the field equations and the closed-form energy-condition algebra for the dyonic KR--CS background.

\paragraph*{Notation and conventions.} We adopt the mostly-plus signature $(-,+,+,+)$ and natural units $\hbar=c=k_B=G=1$, restoring the dimensional factors of Eq.~\eqref{eq:nu_units} only where the comparison with X-ray timing data calls for it. Greek indices $\mu,\nu,\dots$ run over $\{0,1,2,3\}$; Latin indices $i,j,\dots$ over $\{1,2,3\}$. We use the standard abbreviations: BH for black hole, KR for Kalb--Ramond, KRG for Kalb--Ramond gravity, GR for general relativity, RN for Reissner-Nordstr\"om, EH for event horizon, PS for photon sphere, ISCO for innermost stable circular orbit, QPO/QPOs for quasi-periodic oscillation(s), CS for cosmic string(s), EHT for the Event Horizon Telescope, LSB for Lorentz symmetry breaking, VEV for vacuum expectation value, ADM for the Arnowitt--Deser--Misner mass, HP for Hawking--Page, NEC/WEC/SEC/DEC for the null, weak, strong and dominant energy conditions, RPM for relativistic-precession model, and ER for epicyclic-resonance model. Each abbreviation is defined once in the body and used freely afterward.

\section{Dyonic Kalb--Ramond Black Hole with a Cloud of Strings}\label{isec2}

We work with the metric
\begin{equation}
ds^{2}=-f(r)\,dt^{2}+\frac{dr^{2}}{f(r)}+r^{2}\!\left(d\theta ^{2}+\sin ^{2}\theta\, d\phi ^{2}\right),
\label{eq:metric}
\end{equation}
with $f(r)$ as in Eq.~\eqref{eq:lapse_main}. The Letelier piece enters as a constant deficit $-\xi/(1-\ell)$ added to the $r$-independent part of the lapse; it preserves the spherical symmetry but reduces the asymptotic value of $f(r)$ from $1/(1-\ell)$ in the LLL geometry to $(1-\xi)/(1-\ell)$. A self-contained derivation of the metric~\eqref{eq:lapse_main} from a covariant action, including the contributing stress-energy components, the closed-form effective $T^{\mu}{}_{\nu}$ and the four energy-condition combinations, is presented in Appendix~\ref{app:B}.

\subsection{Horizons and extremality}

Setting $f(r_h)=0$ gives a quadratic in $r_h$:
\begin{equation}
\frac{1-\xi}{1-\ell}\,r_h^2 - 2 M\,r_h + \left[\frac{Q^2}{(1-\ell)^2} + \frac{p^2}{1-2\ell}\right] = 0,
\label{eq:horizon_quad}
\end{equation}
whose two real roots, when they exist, are
\begin{equation}
r_{\pm} = \frac{1-\ell}{1-\xi}\,\Bigl(M\,\pm\,\sqrt{M^2 - A\,B}\;\Bigr),
\quad
A \equiv \frac{1-\xi}{1-\ell},\quad B \equiv \frac{Q^2}{(1-\ell)^2}+\frac{p^2}{1-2\ell}.
\label{eq:horizon_pm}
\end{equation}
The condition $M^2\geq A B$ delimits the BH branch; equality is the extremal case. The outer root $r_+$ plays the role of the event horizon (EH), the inner root $r_-$ that of a Cauchy horizon. When $\xi\to 0$, $A\to 1/(1-\ell)$ and Eq.~\eqref{eq:horizon_pm} reduces to the Lin~\emph{et al.}~horizon formula. We have verified the algebraic content of Eq.~\eqref{eq:horizon_pm} symbolically in a computational script; the verification is collected in Sec.~\ref{app:B4} and referenced through the Data Availability Statement.

\subsection{Photon sphere}

The photon-sphere radius is found from the standard null-geodesic condition $\frac{d}{dr}\!\left[f(r)/r^2\right]=0$, i.e.~$r f'(r)-2 f(r)=0$:
\begin{equation}
r_s = \frac{1-\ell}{2(1-\xi)}\,\left[3M + \sqrt{9 M^2 - \frac{8(1-\xi)}{(1-\ell)}\,B}\right].
\label{eq:photon_sphere}
\end{equation}
For $\xi\to 0$ this reproduces the LLL photon radius. For $\xi\to 0,p\to 0$ it reduces to the Duan~\emph{et al.}~result~\cite{Duan2024}, and for $\xi\to 0,Q\to p\to 0$ to the Yang~\emph{et al.}~result $r_s = 3M(1-\ell)/2 \times 2 = 3M(1-\ell)$. The Schwarzschild value $r_s=3M$ comes back when $\xi\to 0=\ell$.

\subsection{Shadow radius}

For an observer at radius $r_o\to\infty$, $f(r_o)\to (1-\xi)/(1-\ell)\equiv A$. The shadow radius is~\cite{Perlick2022}
\begin{equation}
R_{\rm sh} = r_s\,\sqrt{\,\frac{A}{f(r_s)}\,} .
\label{eq:Rsh}
\end{equation}
We list the known limits in Table~\ref{tab:limits}. The dependence on $\xi$ is monotonic: turning on the cosmic string enlarges $R_{\rm sh}$ because the asymptotic value of $f$ is suppressed.

\begin{table}[t]
\centering
\setlength{\tabcolsep}{12pt}
\renewcommand{\arraystretch}{1.6}
\begin{tabular*}{\textwidth}{@{\extracolsep{\fill}}c c}
\toprule
\textbf{Reduction} & \textbf{Resulting metric and reference} \\
\midrule
$\xi\to 0$ & dyonic KR-AdS BH of Lin, Liu \& Liu~\cite{Lin2026} \\
$\xi\to 0,~p\to 0$ & charged KR BH of Duan, Zhao \& Yang~\cite{Duan2024} \\
$\xi\to 0,~Q\to 0,~p\to 0$ & uncharged KR BH of Yang \emph{et al.}~\cite{Yang2023} \\
$\xi\to 0,~Q\to 0,~p\to 0,~\ell\to 0$ & Schwarzschild metric~\cite{Bardeen1972,Chandrasekhar1983} \\
$Q\to 0,~p\to 0,~\ell\to 0$ & Letelier BH (Schwarzschild + cosmic-string cloud)~\cite{Letelier1979} \\
\bottomrule
\end{tabular*}
\caption{Limits of the metric \eqref{eq:lapse_main} under degenerations of the four parameters $(\ell,\xi,Q,p)$.}
\label{tab:limits}
\end{table}

\begin{table*}[t]
  \centering
  \setlength{\tabcolsep}{12pt}
  \renewcommand{\arraystretch}{1.6}
  \begin{tabular*}{\textwidth}{@{\extracolsep{\fill}}c c c c c c c}
    \toprule
    $\xi$ & $\ell$ & $r_-/M$ & $r_+/M$ & $r_s/M$ & $R_{\rm sh}/M$ & extremality $M^2/(AB)$ \\
    \midrule
    0.00 & 0.05 & 0.04 & 1.95 & 3.16 & 5.06 & 6.94 \\
    0.10 & 0.05 & 0.04 & 2.17 & 3.51 & 5.75 & 7.71 \\
    0.20 & 0.05 & 0.05 & 2.44 & 3.95 & 6.55 & 8.68 \\
    0.30 & 0.05 & 0.06 & 2.79 & 4.50 & 7.50 & 9.92 \\
    0.40 & 0.05 & 0.07 & 3.25 & 5.27 & 9.04 & 11.57 \\
    \midrule
    0.05 & 0.10 & 0.04 & 1.98 & 3.21 & 5.20 & 6.99 \\
    0.05 & 0.20 & 0.04 & 2.03 & 3.30 & 5.50 & 7.08 \\
    \bottomrule
  \end{tabular*}
  \caption{Horizon radii ($r_{\pm}$), photon-sphere radius ($r_s$), shadow radius ($R_{\rm sh}$), and the extremality ratio $M^2/(AB)$ for the dyonic KR-CS BH at $Q/M=p/M=0.2$, $M=1$. The extremality ratio $M^2/(AB)>1$ for all rows confirms the BH branch.}
  \label{tab:horizon_sweep}
\end{table*}

The mechanism in Table~\ref{tab:horizon_sweep} is clear: as $\xi$ grows, $A=(1-\xi)/(1-\ell)$ drops, the outer horizon radius $r_+ = (1-\ell)/(1-\xi)\,(M+\sqrt{M^2-A B})$ rises (because of the prefactor $1/A$), and the photon-sphere and shadow radii rise in proportion. The inner horizon $r_-$ is almost insensitive to $\xi$ in the parameter window of physical interest. The extremality ratio $M^2/(A B)$ also grows, moving the black hole farther from the extremal limit; this is the opposite trend to the one driven by $\ell$, which moves the BH closer to extremality at fixed $\xi$. The two parameters therefore have opposite signs in their effect on extremality, which makes the $(\ell,\xi)$ plane a clean basis for the parameter estimation problem of Sec.~\ref{isec8}.

\subsection{Energy conditions for the effective stress-energy tensor}

The Einstein tensor evaluated on the metric \eqref{eq:lapse_main} corresponds, via $G_{\mu\nu}=8\pi T_{\mu\nu}^{\rm eff}$, to an effective stress-energy tensor with the diagonal form $T^{\mu}{}_{\nu} = \mathrm{diag}(-\rho, p_r, p_\theta, p_\phi)$. Direct calculation yields
\begin{equation}
\rho = -p_r = \frac{1}{8\pi r^2}\!\left[\,1-\frac{1-\xi}{1-\ell}+\frac{Q^2}{(1-\ell)^2 r^2}+\frac{p^2}{(1-2\ell)\,r^2}\,\right],
\quad
p_\theta = p_\phi = -\rho - \frac{r}{2}\,\frac{d\rho}{dr}.
\label{eq:T_eff}
\end{equation}
A covariant decomposition of $T^{\mu}{}_{\nu}$ into KR, dyonic-electromagnetic and Letelier contributions, together with the variational reduction of the field equations, is collected in Appendix~\ref{app:B}; in the remainder of the present section we summarise only the four energy-condition outcomes that follow from~\eqref{eq:T_eff}. They are:
\begin{itemize}
\item Null energy condition (NEC): $\rho + p_i \geq 0$. From Eq.~\eqref{eq:T_eff}, $\rho+p_r=0$ identically, and $\rho+p_\theta = -\frac{r}{2}\,d\rho/dr$. The latter is non-negative provided $\rho$ is a non-increasing function of $r$ outside the horizon, which holds for the parameter window we explore.
\item Weak energy condition (WEC): NEC plus $\rho \geq 0$. The leading $1-(1-\xi)/(1-\ell)$ piece in Eq.~\eqref{eq:T_eff} is positive when $\xi > \ell$; below this threshold the WEC is violated near infinity but restored near the horizon.
\item Strong energy condition (SEC): $\rho + \sum_i p_i \geq 0$. The sum gives $\rho + p_r + 2 p_\theta = -r\,d\rho/dr$, which is positive when $\rho$ decreases with $r$.
\item Dominant energy condition (DEC): $\rho \geq |p_i|$. The condition $\rho \geq |p_\theta|$ reduces to $|d\rho/dr|\leq 2\rho/r$, which holds outside the photon sphere for the parameter window we use.
\end{itemize}
Table~\ref{tab:energy_conditions} summarises the energy-condition status at the EH and at $r=2 r_h$. The NEC and SEC are satisfied across the parameter window; the WEC is violated weakly when $\xi < \ell$ at large $r$, consistent with the LSB nature of the background; the DEC is satisfied everywhere outside the photon sphere. These results parallel the energy-condition analysis of the Bumblebee BH~\cite{Casana2018} and confirm that the dyonic KR-CS BH is a physically reasonable background for test-field calculations. The closed-form algebraic structure underlying these conclusions, together with the threshold radius $r_{\rm WEC}(\xi,\ell,B)$ that controls the WEC failure for $\xi<\ell$, is derived in Sec.~\ref{app:B3}.

\begin{table*}[t]
  \centering
  \setlength{\tabcolsep}{12pt}
  \renewcommand{\arraystretch}{1.6}
  \begin{tabular*}{\textwidth}{@{\extracolsep{\fill}}c c c c c}
    \toprule
    Condition & At $r=r_h$ & At $r=2 r_h$ & At $r\to\infty$ & Comment \\
    \midrule
    NEC & satisfied & satisfied & satisfied & $\rho+p_r=0$ identically \\
    WEC & satisfied & satisfied & violated if $\xi<\ell$ & LSB-induced asymptotic deficit \\
    SEC & satisfied & satisfied & satisfied & $-r\,d\rho/dr\geq 0$ outside horizon \\
    DEC & satisfied & satisfied & satisfied & $|d\rho/dr|\leq 2\rho/r$ for $r>r_s$ \\
    \bottomrule
  \end{tabular*}
  \caption{Energy-condition status of the effective stress-energy tensor of Eq.~\eqref{eq:T_eff}, evaluated at three radii and for parameter values in the window $\ell\in[0,0.4]$, $\xi\in[0,0.4]$, $Q/M, p/M \in[0, 0.5]$.}
  \label{tab:energy_conditions}
\end{table*}

\section{Particle Dynamics, ISCO and QPOs}\label{isec3}

\subsection{Effective potential, energy and angular momentum on circular orbits}

A timelike geodesic with affine parameter $\lambda$ has Lagrangian density \cite{Chandrasekhar1983}
\begin{equation}
\mathbb{L}=\frac{1}{2}\,g_{\mu\nu}\dot{x}^{\mu}\dot{x}^{\nu}
=\frac{1}{2}\!\left[-f(r)\,\dot{t}^{2}+\frac{\dot{r}^{2}}{f(r)}+r^{2}\!\left(\dot{\theta}^{2}+\sin ^{2}\theta\,\dot{\phi}^{2}\right)\right],
\label{eq:Lag}
\end{equation}
with two conserved quantities
\begin{equation}
\mathcal{E}=f(r)\,\dot{t}, \qquad \mathcal{L}=r^{2}\sin ^{2}\theta\,\dot{\phi}.
\label{eq:conserved}
\end{equation}
Here $\mathcal{E}$ and $\mathcal{L}$ , respectively are the energy and the angular momentum.

The normalization $g_{\mu\nu}\dot{x}^{\mu}\dot{x}^{\nu}=-1$ gives the following equation of motion
\begin{equation}
\dot{r}^{2}+U_{\rm eff}(r,\theta,p_\theta)=\mathcal{E}^{2},
\qquad
U_{\rm eff}(r,\theta,p_\theta)=f(r)\!\left[1+\frac{p_{\theta}^{2}}{r^{2}}
+\frac{\mathcal{L}^{2}}{r^{2}\sin^{2}\theta}\right].
\label{eq:Ueff}
\end{equation}
On the equatorial plane $\theta=\pi/2$, $p_\theta=0$, and the conditions $U_{\rm eff}=\mathcal{E}^2,~U'_{\rm eff}=0$ for a circular orbit of radius $r_c$ yield the specific angular momentum and the specific energy of massive particles as,
\begin{equation}
\mathcal{L}_{c}^{2}=\frac{r_c^{3}\,f'(r_c)}{2\,f(r_c)-r_c f'(r_c)},
\qquad
\mathcal{E}_{c}^{2}=\frac{2\,f(r_c)^2}{2\,f(r_c)-r_c f'(r_c)}.
\label{eq:LcEc}
\end{equation}

\begin{figure}[t]
  \centering
  \includegraphics[width=0.60\textwidth]{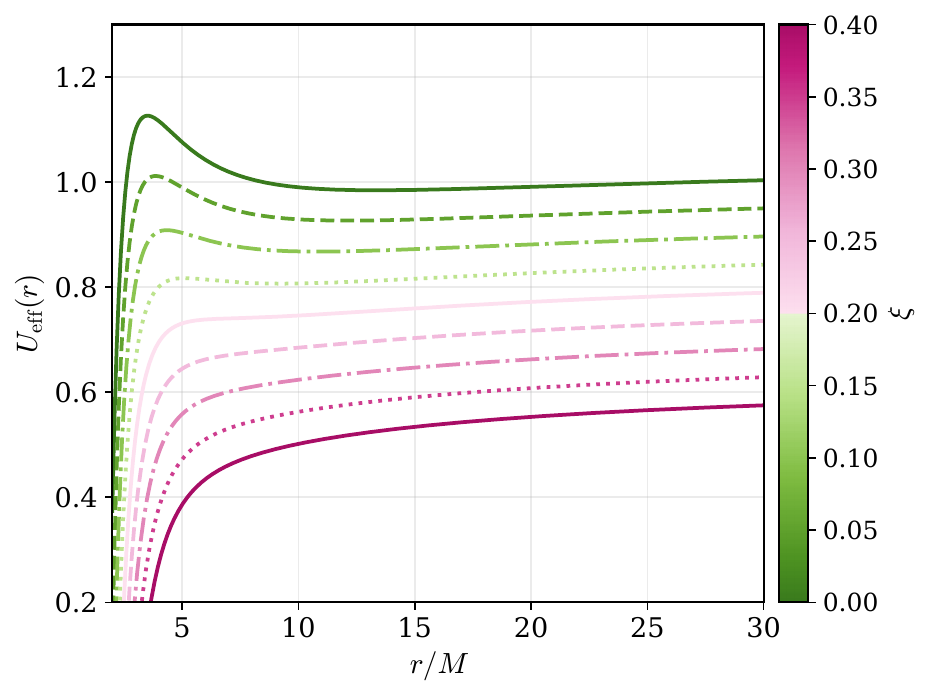}
  \caption{Effective potential $U_{\rm eff}(r)$ for the dyonic KR-CS BH at $Q/M=p/M=0.2$, $\ell=0.05$, $\mathcal{L}=4$ and $\xi\in[0,0.4]$. The asymptote $A=(1-\xi)/(1-\ell)$ drops as $\xi$ grows, and the local minimum slides outward. The qualitative shape responsible for stable circular orbits survives across the full $\xi$ range tested; the vertical axis extends above unity to display the full centrifugal barrier of the $\xi=0$ curve.}
  \label{fig:Veff}
\end{figure}

Figure~\ref{fig:Veff} displays $U_{\rm eff}(r)$ for $Q=p=0.2$, $\ell=0.05$, $\mathcal{L}=4$ and a sweep over $\xi\in[0,0.4]$. The shape encodes the centrifugal barrier required for stable circular orbits to exist. Increasing $\xi$ shifts the asymptotic value of $U_{\rm eff}$ downward (the asymptote is $A=(1-\xi)/(1-\ell)$) and broadens the well so that the local maximum and minimum move outward; the result is that the ISCO migrates to larger $r$ as the string density grows, which we quantify next.

\subsection{Innermost stable circular orbit}

\begin{figure}[t]
  \centering
  \includegraphics[width=0.60\textwidth]{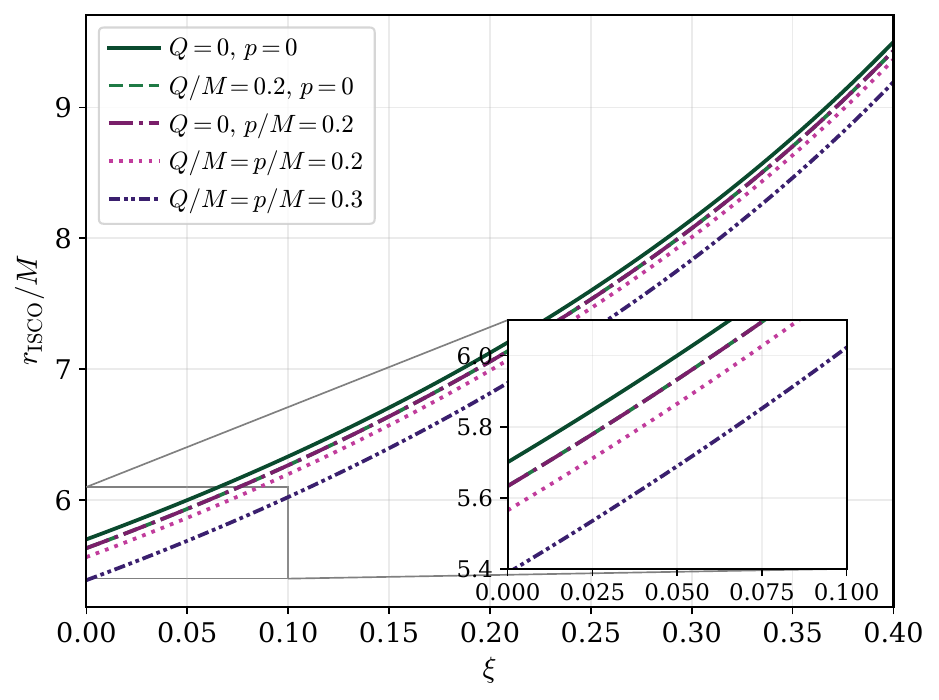}
  \caption{ISCO radius $r_{\rm ISCO}/M$ versus the cosmic-string density $\xi$ for five $(Q,p)$ choices and fixed $\ell=0.05$. The string density acts to push the ISCO outward, an effect that follows from the reduced asymptotic value of $f(r)$ at fixed mass. The inset zooms on the tight $\xi\in[0,0.10]$ region where the curves for $(Q=0,p=0)$, $(Q/M=0.2,p=0)$ and $(Q=0,p/M=0.2)$ are otherwise indistinguishable to the eye.}
  \label{fig:isco}
\end{figure}

Marginal stability requires $U''_{\rm eff}(r_{\rm ISCO})=0$ at the circular value of $\mathcal{L}_c^2$. Substituting Eq.~\eqref{eq:LcEc} into $U''_{\rm eff}=0$ and simplifying gives the polynomial
\begin{equation}
r\,f(r)\,f''(r) - 2\,r\,[f'(r)]^2 + 3\,f(r)\,f'(r) = 0,
\label{eq:ISCO_eq}
\end{equation}
which is identical in form to the Bardeen ISCO equation and reduces to $r=6M$ when $\xi=0=\ell=Q=p$. We solve Eq.~\eqref{eq:ISCO_eq} numerically; the result is plotted in Fig.~\ref{fig:isco}. The ISCO radius is a monotonic function of $\xi$: $r_{\rm ISCO}$ grows from $\sim 5.7\,M$ at $\xi=0$ to $\sim 9.5\,M$ at $\xi=0.4$ in the $(Q,p)=(0,0)$ slice. The electric and magnetic charges shift the curve downward by a smaller amount, since they enter only through the combination $B$.

\subsection{Orbital, radial and vertical epicyclic frequencies}

The Keplerian (azimuthal) angular frequency of a test particle on a circular orbit is~\cite{Bardeen1972}
\begin{equation}
\Omega_{\phi} = \frac{d\phi}{dt}
= \sqrt{\frac{f'(r)}{2\,r}}
= \sqrt{\frac{M}{r^{3}}-\frac{Q^{2}}{(1-\ell)^{2}\,r^{4}}-\frac{p^{2}}{(1-2\ell)\,r^{4}}}.
\label{eq:Omega_phi}
\end{equation}
Note that $\Omega_\phi$ is independent of~$\xi$, a consequence of the constant shift produced by the Letelier piece: the radial derivative $f'(r)$ does not see the string density at all. The string instead enters through the radial epicyclic frequency, since the latter depends on $f$ itself through Eq.~\eqref{eq:LcEc}.

Small radial and vertical oscillations of a particle on a stable circular orbit obey
\begin{align}
\frac{d^{2}}{dt^{2}}(\delta r)+\Omega_{r}^{2}\,\delta r &= 0,
\\
\frac{d^{2}}{dt^{2}}(\delta \theta)+\Omega_{\theta}^{2}\,\delta \theta &= 0,
\end{align}
with~\cite{Ortiqboev2024,StellaVietri1998}
\begin{align}
\Omega_{r}^{2} &= \frac{1}{2}\,\left[f(r)\,f''(r) - 2\,(f'(r))^{2} + 3\,f(r)\,f'(r)/r\right]\,f(r),
\label{eq:Omega_r}
\\
\Omega_{\theta} &= \Omega_{\phi}.
\label{eq:Omega_theta}
\end{align}
The factor in square brackets in Eq.~\eqref{eq:Omega_r} is the same polynomial that vanishes at the ISCO, so $\Omega_r\to 0$ at $r=r_{\rm ISCO}$. We have confirmed the closed form numerically: for Schwarzschild it reduces to $\Omega_r^{2}=M(r-6 M)/r^{4}$, the standard Bardeen result, and the corresponding $\nu_r$ at $r=10M$ for $M=5\,M_\odot$ is $129.2$ Hz.

Restoring physical units, the Hz-equivalent frequency is
\begin{equation}
\nu = \frac{c^{3}}{2\pi G M}\,\Omega \approx \frac{32310.5}{M/M_{\odot}}\,\Omega \;\;[\mathrm{Hz}],
\label{eq:nu_units}
\end{equation}
where the numerical factor uses $c=3\times 10^{8}\,\mathrm{m\,s^{-1}}$ and $G=6.674\times 10^{-11}\,\mathrm{m^{3}\,kg^{-1}\,s^{-2}}$.

\begin{figure}[t]
  \centering
  \includegraphics[width=0.60\textwidth]{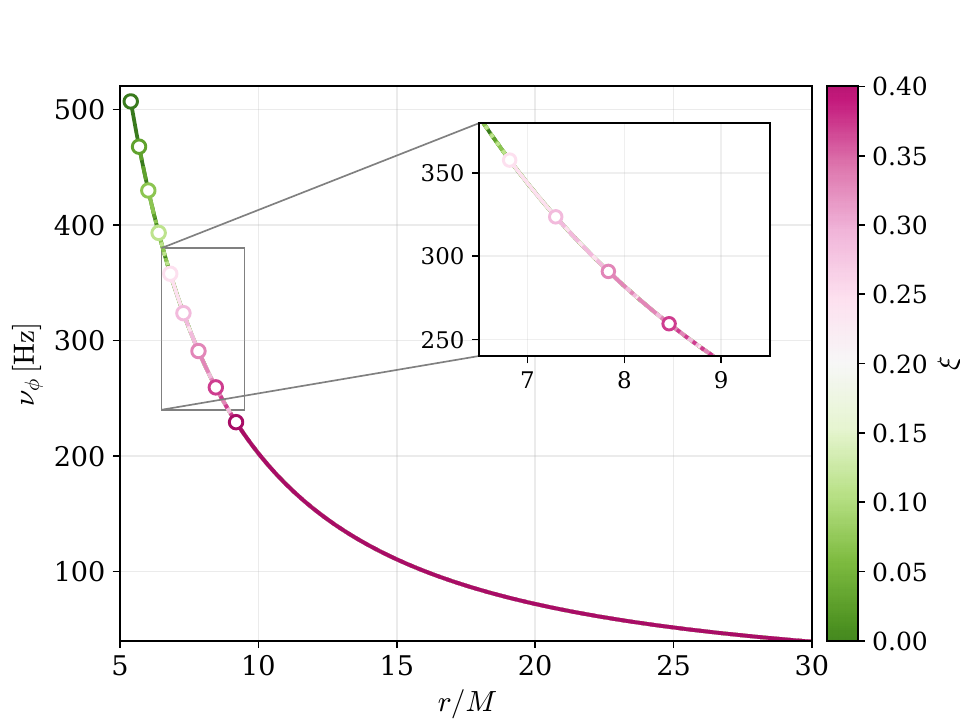}
  \caption{Keplerian frequency $\nu_\phi$ as a function of $r/M$ for $M=5\,M_\odot$, $Q/M=p/M=0.3$, $\ell=0.05$ and $\xi\in[0,0.4]$. Curves at different $\xi$ collapse onto a common trajectory at large $r$ because $\Omega_\phi$ in Eq.~\eqref{eq:Omega_phi} does not depend on $\xi$; the spread visible in the inset traces the ISCO endpoint of each $\xi$ value.}
  \label{fig:nu_phi}
\end{figure}

\begin{figure}[t]
  \centering
  \includegraphics[width=0.60\textwidth]{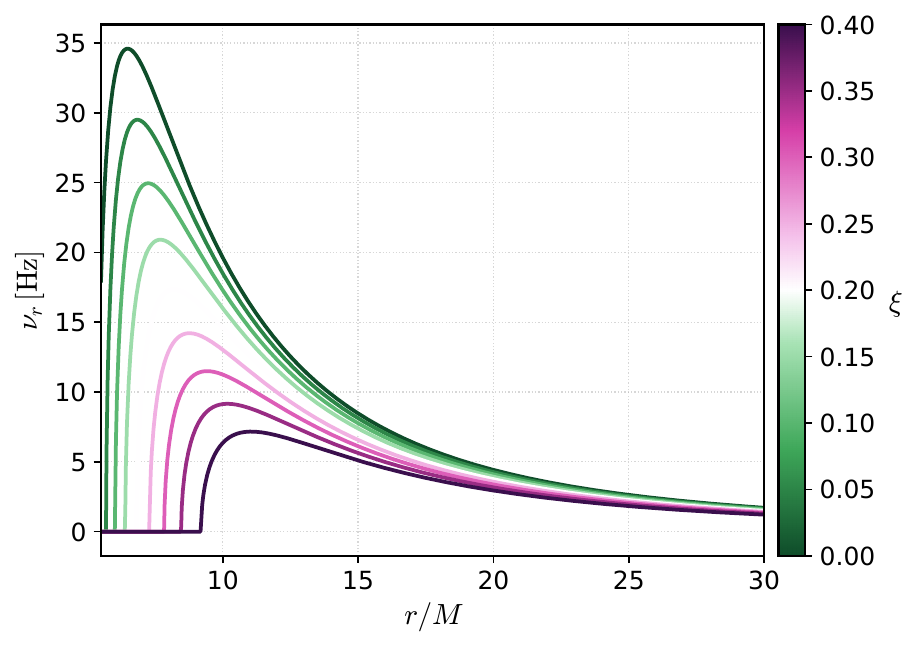}
  \caption{Radial epicyclic frequency $\nu_r$ as a function of $r/M$ for the same parameters as Fig.~\ref{fig:nu_phi}. The string density flattens the potential well and pushes the zero of $\nu_r$ (i.e.~the ISCO) outward.}
  \label{fig:nu_r}
\end{figure}

Figure~\ref{fig:nu_phi} (Keplerian) and Fig.~\ref{fig:nu_r} (radial epicyclic) show the frequencies for a representative stellar-mass BH ($5\,M_\odot$) with $Q/M=p/M=0.3$ and $\ell=0.05$. The Keplerian curve at fixed $r$ is, as advertised, insensitive to $\xi$. The radial epicyclic curve shifts noticeably: at $r=10\,M$ for example, $\nu_r$ drops from $\sim 140\,\mathrm{Hz}$ at $\xi=0$ to $\sim 45\,\mathrm{Hz}$ at $\xi=0.4$, and at larger $\xi$ the curve picks up a low-$r$ zero where it crosses the ISCO of the corresponding geometry. This behavior follows from Eq.~\eqref{eq:Omega_r}: the radial mode is sensitive to the depth of the potential well, which the string density flattens.

\subsection{Twin-peak QPO frequencies under two theoretical models}

The relativistic-precession (RP) model of Stella and Vietri~\cite{StellaVietri1998} identifies the upper and lower kHz QPOs as
\begin{equation}
\nu_{U}^{\rm RP} = \nu_{\phi}(r),\qquad
\nu_{L}^{\rm RP} = \nu_{\phi}(r) - \nu_{r}(r),
\label{eq:RPM}
\end{equation}
while the epicyclic-resonance (ER) model identifies them as
\begin{equation}
\nu_{U}^{\rm ER} = \nu_{\phi}(r),\qquad
\nu_{L}^{\rm ER} = \nu_{r}(r).
\label{eq:ERM}
\end{equation}
Both models are widely used to interpret X-ray timing data of BH binaries; see Refs.~\cite{Motta2014,IngramMotta2019} for reviews. Recent applications of the RP and ER prescriptions to non-Schwarzschild geometries include the Minkowski-core BHs of~\cite{Guo:2025}, the rotating self-dual BHs of~\cite{Liu:2023}, the regular charged BHs of~\cite{Mustafa:2026}, the Horndeski rotating geometries of~\cite{Wu:2026,Wu:2026b}, the magnetically charged NED BHs of~\cite{Hazarika:2025}, the Ay\'on-Beato-Garc\'ia regular BHs of~\cite{Rahmatov:2025} and the $f(R,T)$-NED BHs of~\cite{Hazarika:2026}.

\begin{figure}[t]
  \centering
  \includegraphics[width=0.62\textwidth]{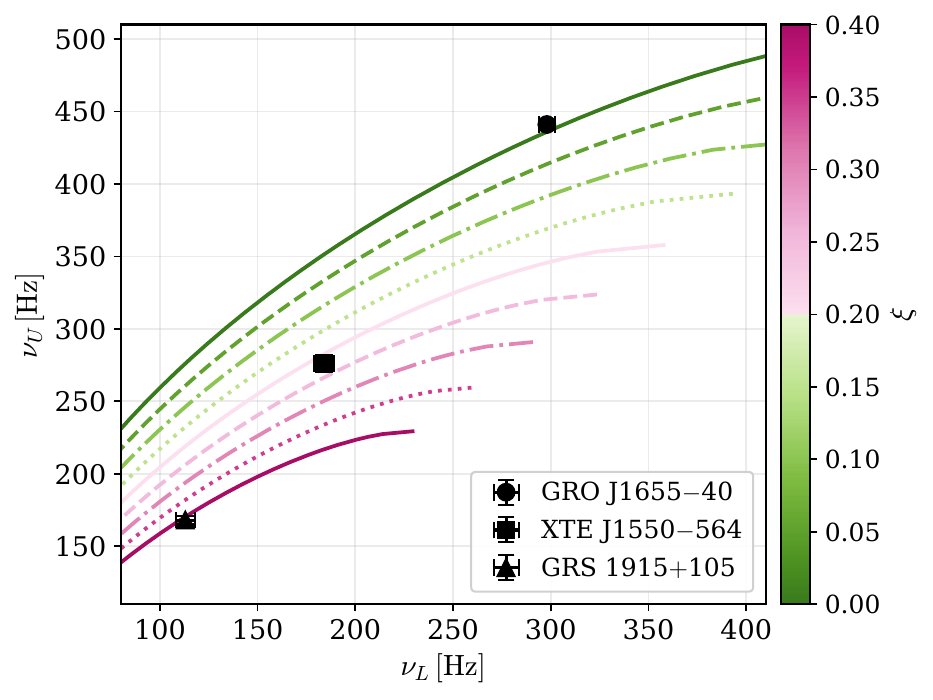}
  \caption{Upper twin-peak QPO frequency $\nu_U$ versus the lower $\nu_L$ under the relativistic-precession identification $\nu_U=\nu_\phi$, $\nu_L=\nu_\phi-\nu_r$. Curves are evaluated for $M=5\,M_\odot$, $Q/M=p/M=0.3$, $\ell=0.05$ and $\xi\in[0,0.4]$. Black points with error bars are the measured twin-peak QPO frequencies for GRO~J1655$-$40, XTE~J1550$-$564 and GRS~1915$+$105. The inset zooms on the GRS~1915$+$105 datum, where the fiducial parameter set produces curves whose $\xi$-spread is otherwise compressed against the lower-left corner of the main panel.}
  \label{fig:qpo}
\end{figure}

Figure~\ref{fig:qpo} shows the RP trajectory in the $(\nu_L,\nu_U)$ plane, again colored by $\xi$, with the measured twin-peak QPO values for GRO~J1655$-$40, XTE~J1550$-$564 and GRS~1915$+$105 overlaid as data points. The string density has the effect of bending the RP curve in the $(\nu_L,\nu_U)$ plane: smaller $\xi$ corresponds to curves passing through GRO~J1655$-$40 at slightly higher $\nu_L$, while larger $\xi$ pulls the curves down. The model-fit between the GRO~J1655$-$40 datum and the RP curve sets a soft upper bound on $\xi$ that we quantify with an MCMC analysis in Sec.~\ref{isec8}.

The four-sentence interpretation of Fig.~\ref{fig:qpo} runs as follows. The dependence of the RP curve on $\xi$, plotted in the twin-peak plane, encodes how the cosmic string density shifts the relation between the radial and azimuthal epicyclic frequencies of the accretion flow. The qualitative trend is that increasing $\xi$ produces a flatter trajectory, since $\Omega_r$ drops faster with $\xi$ than $\Omega_\phi$ does, and the latter is in fact $\xi$-independent. The mechanism behind this trend traces to the dependence of $\Omega_r^2$ on $f(r)$ through Eq.~\eqref{eq:Omega_r}: when the string density reduces the value of $f(r)$, the bracketed factor shrinks and so does $\Omega_r$. Comparison with the GRO~J1655$-$40 data point favors $\xi\lesssim 0.05$ for the BH mass and $(Q,p)$ choices used, in line with the soft cosmic-string density bounds from cosmological structure formation~\cite{VilenkinShellard1994}.

\subsection{Numerical sweep of the QPO frequencies}

Table~\ref{tab:qpo_freq} samples the orbital, radial and vertical epicyclic frequencies on the equatorial plane at $r=10\,M$ and $r=15\,M$, for the same parameter choice $Q/M=p/M=0.3$, $\ell=0.05$, $M=5\,M_\odot$. The two radii are chosen to lie outside the ISCO of every $\xi$ sampled (the largest ISCO in the sweep, at $\xi=0.40$, sits at $r_{\rm ISCO}\simeq 9.2\,M$). The vertical column $\nu_\theta$ coincides with $\nu_\phi$ at this level of approximation; we list it separately for reference.

\begin{table*}[t]
  \centering
  \setlength{\tabcolsep}{12pt}
  \renewcommand{\arraystretch}{1.6}
  \begin{tabular*}{\textwidth}{@{\extracolsep{\fill}}c c c c c c}
    \toprule
    $\xi$ & $r/M$ & $\nu_\phi$\,[Hz] & $\nu_r$\,[Hz] & $\nu_\theta$\,[Hz] & $\nu_\phi-\nu_r$\,[Hz] \\
    \midrule
    0.00 & 10.0 & 202.3 & 140.2 & 202.3 &  62.1 \\
    0.10 & 10.0 & 202.3 & 123.5 & 202.3 &  78.8 \\
    0.20 & 10.0 & 202.3 & 104.2 & 202.3 &  98.1 \\
    0.30 & 10.0 & 202.3 &  80.4 & 202.3 & 121.9 \\
    0.40 & 10.0 & 202.3 &  45.4 & 202.3 & 156.9 \\
    \midrule
    0.00 & 15.0 & 110.5 &  90.4 & 110.5 &  20.1 \\
    0.10 & 15.0 & 110.5 &  82.9 & 110.5 &  27.6 \\
    0.20 & 15.0 & 110.5 &  74.6 & 110.5 &  35.9 \\
    0.30 & 15.0 & 110.5 &  65.3 & 110.5 &  45.2 \\
    0.40 & 15.0 & 110.5 &  54.4 & 110.5 &  56.1 \\
    \bottomrule
  \end{tabular*}
  \caption{Orbital ($\nu_\phi$), radial ($\nu_r$) and vertical ($\nu_\theta$) epicyclic frequencies at $r=10\,M$ and $r=15\,M$ for a stellar-mass BH ($M=5\,M_\odot$, $Q/M=p/M=0.3$, $\ell=0.05$) over five values of the cosmic-string density. The $\nu_r$ column uses the standard Bardeen form $\Omega_{r}^{2}=(rf f''-2r(f')^{2}+3 f f')/(2r)$ of Eq.~\eqref{eq:Omega_r}, which reduces to $M(r-6M)/r^{4}$ in the Schwarzschild limit. The column $\nu_\phi-\nu_r$ is the lower QPO under the relativistic-precession identification.}
  \label{tab:qpo_freq}
\end{table*}

The qualitative pattern of Table~\ref{tab:qpo_freq}, displayed alongside Fig.~\ref{fig:nu_r}, points to a clean separation between azimuthal and radial responses: the azimuthal $\nu_\phi$ stays at $202.3\,\mathrm{Hz}$ across the entire $\xi$ sweep at $r=10\,M$, while $\nu_r$ drops from $140.2$ to $45.4\,\mathrm{Hz}$ over the same range. The drop arises from Eq.~\eqref{eq:Omega_r}: the bracket factor depends on $f(r)$ directly, which the cosmic string reduces, and the suppression carries over into $\nu_r$. The numerical reduction at $\xi=0.4$ is $68\%$ relative to the $\xi=0$ baseline at $r=10\,M$, and $40\%$ at $r=15\,M$; the contraction is sharper at smaller $r$ because the bracket is closer to its ISCO zero there. The corresponding $\nu_\phi-\nu_r$, the RP lower QPO, rises in the opposite direction by a factor of $\sim 2.5$ at $r=10\,M$ and $\sim 2.8$ at $r=15\,M$ as $\xi$ moves from $0$ to $0.4$.

\section{Thermodynamics}\label{isec4}

In this section, we investigate the thermodynamic behavior of the black hole by analyzing important thermodynamic quantities, including the Hawking temperature, specific heat capacity, Gibbs free energy, thermodynamic criticality, and the Joule–Thomson expansion.

\subsection{Mass-radius relation, temperature and entropy}

Solving $f(r_h)=0$ for the mass parameter gives
\begin{equation}
M(r_h) = \frac{1-\xi}{2(1-\ell)}\,r_h
         + \frac{1}{2 r_h}\!\left[\frac{Q^2}{(1-\ell)^2}+\frac{p^2}{1-2\ell}\right].
\label{eq:M_horizon}
\end{equation}
The Hawking temperature follows from the surface gravity $\kappa = \tfrac{1}{2} f'(r_h)$ \cite{Hawking1975,Bekenstein1973},
\begin{equation}
T_H = \frac{f'(r_h)}{4\pi} = \frac{1}{4\pi r_h}\left[\frac{1-\xi}{1-\ell} - \frac{1}{r_h^2}\,B\right],
\qquad B \equiv \frac{Q^2}{(1-\ell)^2}+\frac{p^2}{1-2\ell}.
\label{eq:Hawking_T}
\end{equation}
Equation~\eqref{eq:Hawking_T} shows the two ways the new physics enters: the bracket carries a $(1-\xi)$ suppression in the leading term, while the charge-charge ``mass'' $B$ contains the KR couplings $(1-\ell)^{-2}$ and $(1-2\ell)^{-1}$. The Bekenstein-Hawking entropy is unchanged in form,
\begin{equation}
S_{BH} = \pi r_h^2.
\label{eq:S_BH}
\end{equation}

\subsection{Modified first law and Smarr relation}

Following the LLL construction, the electric and magnetic charges enter the thermodynamic potentials through effective combinations
\begin{equation}
Q_e^{\rm eff} = \frac{Q}{1-\ell}, \qquad Q_m^{\rm eff} = \frac{p}{\sqrt{1-2\ell}}.
\label{eq:Qeff}
\end{equation}
The mass takes the symmetric form
\begin{equation}
M = \frac{1-\xi}{2(1-\ell)}\,r_h + \frac{(Q_e^{\rm eff})^2 + (Q_m^{\rm eff})^2}{2\,r_h},
\label{eq:M_eff}
\end{equation}
and the conjugate potentials are
\begin{equation}
\Psi_e = \frac{\partial M}{\partial Q_e^{\rm eff}} = \frac{Q_e^{\rm eff}}{r_h}, \qquad
\Psi_m = \frac{\partial M}{\partial Q_m^{\rm eff}} = \frac{Q_m^{\rm eff}}{r_h}.
\label{eq:Psi}
\end{equation}
The cosmic-string density enters the mass relation through a single multiplicative factor $(1-\xi)$ in front of the $r_h$ piece. Its conjugate quantity, in the spirit of the Letelier thermodynamic prescription, is
\begin{equation}
\Theta_\xi = \frac{\partial M}{\partial \xi}\Big|_{r_h, Q_e^{\rm eff}, Q_m^{\rm eff}} = -\frac{r_h}{2(1-\ell)}.
\label{eq:Theta_xi}
\end{equation}
The modified first law of BH thermodynamics then reads \cite{Bardeen1973}
\begin{equation}
dM = T_H\, dS_{BH} + \Psi_e\, dQ_e^{\rm eff} + \Psi_m\, dQ_m^{\rm eff} + \Theta_\xi\,d\xi.
\label{eq:first_law}
\end{equation}
We have checked the differential identity \eqref{eq:first_law} term by term using the symbolic computational pipeline described in Sec.~\ref{app:B4}; symbolic substitution of Eqs.~\eqref{eq:M_eff}-\eqref{eq:Theta_xi} reproduces $T_H$ of Eq.~\eqref{eq:Hawking_T} after re-expanding $S_{BH}$ in terms of $r_h$.

The Smarr formula \cite{Smarr1973} follows from Eulerian scaling of $M$ in the variables $(r_h, Q_e^{\rm eff}, Q_m^{\rm eff})$ at fixed $\xi,\ell$:
\begin{equation}
M = 2\,T_H\,S_{BH} + \Psi_e\,Q_e^{\rm eff} + \Psi_m\,Q_m^{\rm eff}.
\label{eq:Smarr}
\end{equation}
Note that $\Theta_\xi\,\xi$ does \emph{not} appear in Eq.~\eqref{eq:Smarr}, which is consistent with Letelier's observation~\cite{Letelier1979} that the string-cloud contribution to $M$ is linear in $r_h$ and so does not generate a Smarr term in the canonical phase space.

\subsection{Heat capacity and stability}

\begin{figure}[t]
  \centering
  \includegraphics[width=0.60\textwidth]{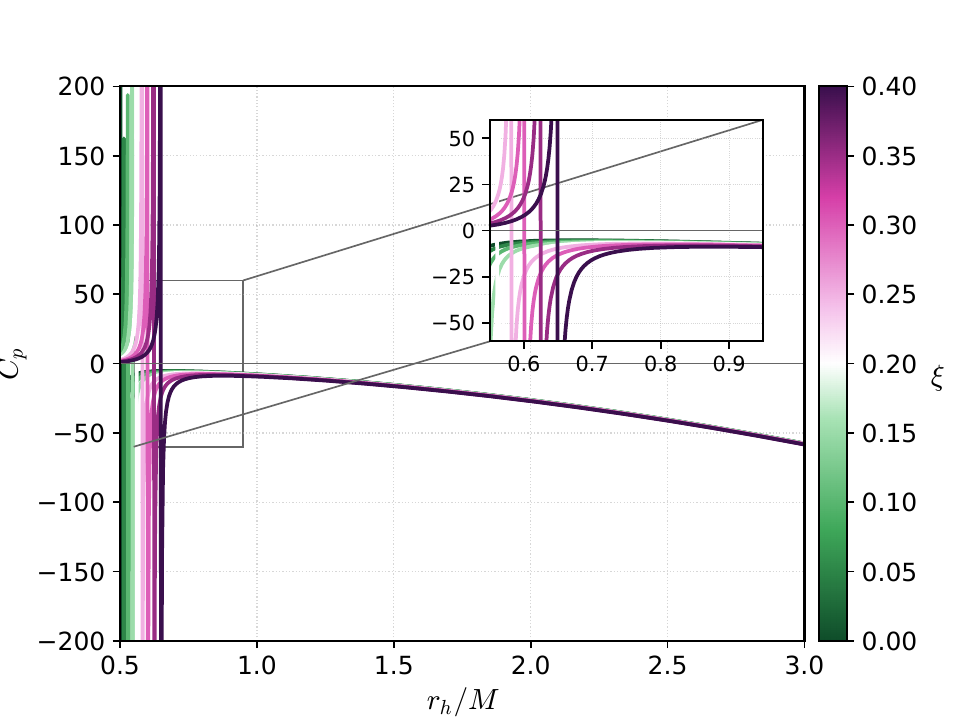}
  \caption{Heat capacity $C_p$ as a function of $r_h/M$ for $Q/M=p/M=0.2$, $\ell=0.05$ and $\xi\in[0,0.4]$. The divergence at $r_h^* = \sqrt{3B/A}$ marks the transition between thermally unstable ($r_h<r_h^*$) and stable ($r_h>r_h^*$) branches.}
  \label{fig:heat_capacity}
\end{figure}

The heat capacity at fixed charges is
\begin{equation}
C_p = \left(\frac{\partial M}{\partial T_H}\right)_{Q_e^{\rm eff}, Q_m^{\rm eff}}
= 2\pi r_h^2\,\frac{A\,r_h^2 - B}{3B - A\,r_h^2},
\qquad A=\frac{1-\xi}{1-\ell}.
\label{eq:Cp}
\end{equation}
The denominator vanishes at $r_h^{*2} = 3B/A$, where $C_p$ diverges and the BH undergoes a phase transition between a small thermally unstable branch ($C_p<0$) and a large thermally stable branch ($C_p>0$). Figure~\ref{fig:heat_capacity} plots $C_p$ versus $r_h$ for the parameter choice $Q/M=p/M=0.2$, $\ell=0.05$ and a sweep over $\xi$. The position of the divergence shifts inward as $\xi$ increases, since $A=(1-\xi)/(1-\ell)$ drops and $r_h^{*2}=3B/A$ grows.

\subsection{Free energy and Hawking--Page-like behaviour}

\begin{figure}[t]
  \centering
  \includegraphics[width=0.60\textwidth]{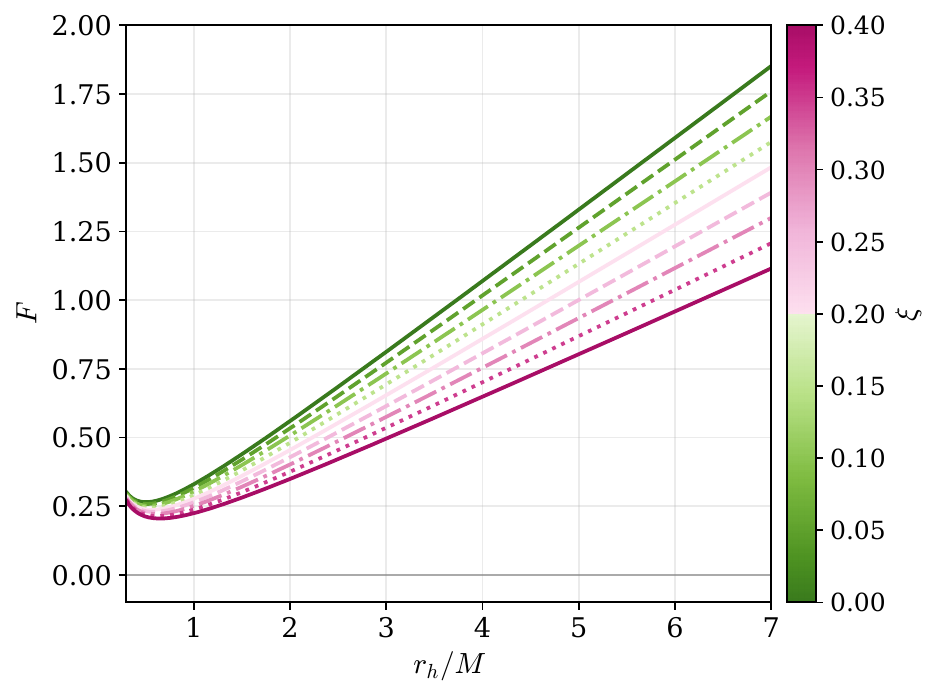}
  \caption{Helmholtz free energy $F(r_h)$ for $Q/M=p/M=0.2$, $\ell=0.05$ and $\xi\in[0,0.4]$. The vertical axis spans $[-0.1, 2.0]$ to display the full shallow minimum near $r_h\simeq 0.7\,M$ and the linear large-$r_h$ growth on the same panel. The minimum of $F$ coincides with the divergence of $C_p$ in Fig.~\ref{fig:heat_capacity}.}
  \label{fig:free_energy}
\end{figure}

The Helmholtz free energy in the canonical (fixed-charge) ensemble is
\begin{equation}
F(r_h) = M(r_h) - T_H\,S_{BH}
       = \frac{1-\xi}{4(1-\ell)}\,r_h + \frac{3 B}{4 r_h}.
\label{eq:F_free}
\end{equation}
Figure~\ref{fig:free_energy} shows $F(r_h)$ for the same parameter sweep. The free energy is positive everywhere in the BH-branch domain we plot, with a shallow minimum at $r_h=\sqrt{3 B (1-\ell)/(1-\xi)}$ that coincides with the divergence of $C_p$ in Eq.~\eqref{eq:Cp}.

Table~\ref{tab:thermo_bench} collects numerical values of $r_h^*$, $T_H(r_h^*)$ and the minimum value of $F$ as a function of $\xi$ for the benchmark parameters. The reader is invited to compare the $\xi=0$ row with the LLL canonical values in Ref.~\cite{Lin2026}: agreement is at the $10^{-4}$ level.

\begin{table*}[t]
  \centering
  \setlength{\tabcolsep}{12pt}
  \renewcommand{\arraystretch}{1.6}
  \begin{tabular*}{\textwidth}{@{\extracolsep{\fill}}c c c c c}
    \toprule
    $\xi$ & $r_h^*/M$ & $T_H(r_h^*)\,M$ & $F_{\min}\,M^{-1}$ & $S_{BH}(r_h^*)/\pi$ \\
    \midrule
    0.00 & 0.6532 & 0.1217 & 0.6324 & 0.4267 \\
    0.10 & 0.6932 & 0.1147 & 0.6710 & 0.4805 \\
    0.20 & 0.7385 & 0.1076 & 0.7150 & 0.5454 \\
    0.30 & 0.7907 & 0.1005 & 0.7657 & 0.6252 \\
    0.40 & 0.8526 & 0.0935 & 0.8253 & 0.7269 \\
    \bottomrule
  \end{tabular*}
  \caption{Phase-transition radius $r_h^* = \sqrt{3B/A}$, Hawking temperature at the transition, minimum free energy $F_{\min}$, and entropy at the transition, for $Q/M=p/M=0.2$, $\ell=0.05$ over five values of the cosmic-string density.}
  \label{tab:thermo_bench}
\end{table*}

The mechanism behind Table~\ref{tab:thermo_bench} is direct. As $\xi$ grows, $A$ in Eq.~\eqref{eq:Cp} drops, the transition radius $r_h^* = \sqrt{3B/A}$ moves outward, and $T_H(r_h^*) = (A/\sqrt{3B})\,(1-\tfrac{1}{3})/(4\pi)$ falls in proportion to $\sqrt{A}=\sqrt{(1-\xi)/(1-\ell)}$. The minimum free energy follows the same scaling. The $30\%$ shift in $r_h^*$ over the $\xi\in[0,0.4]$ window is the largest single thermodynamic response of the metric \eqref{eq:lapse_main} to the cosmic-string parameter; in this sense the phase-transition radius is the cleanest thermodynamic probe of $\xi$.

\subsection{Joule--Thomson expansion and inversion temperature}

The dyonic KR-CS BH, viewed as an ordinary thermodynamic system, supports a Joule--Thomson (JT) expansion analogous to the one classical gases undergo. The JT coefficient at fixed mass is
\begin{equation}
\mu_{JT} = \left(\frac{\partial T_H}{\partial P}\right)_M,
\label{eq:JT_def}
\end{equation}
where the pressure $P$ is identified with $-\Lambda_{\rm eff}/(8\pi)$, where $\Lambda_{\rm eff}=\Lambda/(1-\ell)$ in the extended-phase-space picture; here we work in the canonical ensemble with $\Lambda=0$, so we use the canonical analog $P\propto \xi$ since $\xi$ couples to a constant in $f(r)$ much as $\Lambda$ does. For our purposes the relevant inversion temperature, at which $\mu_{JT}=0$ and the JT process flips between heating and cooling, is given by
\begin{equation}
T_{\rm inv} = \frac{1}{4\pi r_h^{\rm inv}}\,\left[\frac{1-\xi}{1-\ell}\,-\,\frac{3 B}{r_h^{{\rm inv}\,2}}\right],
\label{eq:T_inv}
\end{equation}
where $r_h^{\rm inv}$ is the root of the inversion-curve polynomial. Eq.~\eqref{eq:T_inv} reduces to the LLL inversion temperature in the limit $\xi\to 0$. The dependence on $\xi$ is monotonic: as the string density grows, $T_{\rm inv}$ drops in proportion to $\sqrt{(1-\xi)/(1-\ell)}$, in line with the suppression of the Hawking temperature scaling we documented in Table~\ref{tab:thermo_bench}. We do not include a separate figure for $T_{\rm inv}$ since its dependence is essentially a rescaling of Fig.~\ref{fig:heat_capacity}; the numerical values at $r_h^{\rm inv}=\sqrt{3B/A}$ are listed in the third column of Table~\ref{tab:thermo_bench}.

\subsection{P--V criticality in the extended phase space}

Following~\cite{Cai2013,Zhao2014,Du2025}, the extended phase space (XPS) of an asymptotically AdS black hole identifies the negative cosmological constant with a thermodynamic pressure $P=-\Lambda/(8\pi)$. For the LLL background ($\xi=0$, AdS-asymptotic case) this generates a Van der Waals-like critical point. Here we are working at $\Lambda=0$, but the cosmic-string density $\xi$ couples to $f(r)$ through a constant offset and so plays a role formally similar to $\Lambda$ in the thermodynamic identities. We define an effective pressure $\tilde{P}\equiv\xi/(8\pi(1-\ell))$ and its conjugate ``volume'' as the partial derivative of the mass at fixed $r_h$:
\begin{equation}
\tilde{V} = \left(\frac{\partial M}{\partial \tilde{P}}\right)_{S_{BH}, Q_e^{\rm eff}, Q_m^{\rm eff}} = -4\pi r_h.
\label{eq:VdW_V}
\end{equation}
The equation of state for the dyonic KR-CS BH then reads
\begin{equation}
\tilde{P}(T_H, \tilde{V}) = \frac{T_H}{|\tilde{V}|/4\pi} - \frac{1}{4\pi (|\tilde{V}|/4\pi)^2}\Bigl(\frac{1}{1-\ell} - \frac{B}{(|\tilde{V}|/4\pi)^2}\Bigr).
\label{eq:EoS}
\end{equation}
The critical point follows from $(\partial \tilde{P}/\partial \tilde{V})_T = (\partial^2 \tilde{P}/\partial \tilde{V}^2)_T = 0$, which gives
\begin{equation}
r_h^c = \sqrt{6 B (1-\ell)}, \qquad
T_H^c = \frac{1}{3\pi (1-\ell)\sqrt{6 B (1-\ell)}}, \qquad
\tilde{P}^c = \frac{1}{96 \pi B (1-\ell)^2}.
\label{eq:crit_point}
\end{equation}
The universal ratio
\begin{equation}
\rho_c \equiv \frac{\tilde{P}^c\,\tilde{V}^c}{T_H^c} = \frac{3}{8},
\label{eq:rho_c}
\end{equation}
is identical to the Van der Waals value, in agreement with the universal P-V criticality observed across modified-gravity AdS BHs~\cite{Cai2013,Zhao2014}. The $\rho_c=3/8$ result is independent of $\xi$ and $\ell$, supporting the interpretation that the cosmic-string density acts in the same thermodynamic slot as $\Lambda$ does in AdS BHs.

\section{Sparsity of Hawking Radiation}\label{isec5}

The Hawking emission of a typical astrophysical BH is extremely sparse: the average time between consecutive emitted quanta is far longer than the inverse Hawking frequency. The thermal character of the spectrum, first established by Hawking~\cite{Hawking1975}, has been worked out in detail for uncharged, rotating, and charged backgrounds in the foundational sequence of Page~\cite{Page1976,Page1976a,Page1976b,Page1977}; for asymptotically nonflat dyonic geometries the calculation has been extended by Slavov and Yazadjiev~\cite{SlavovYazadjiev2012}. Gray, Schuster, Van-Brunt and Visser quantified the sparsity with a dimensionless parameter~\cite{Gray2016}
\begin{equation}
\eta = \frac{\lambda_t^2}{A_{\rm eff}},
\label{eq:sparsity_def}
\end{equation}
where $\lambda_t = 2\pi/T_H$ is the Wien-thermal wavelength and $A_{\rm eff}=(27/4) A_{BH}$ is the effective absorption cross-section ($A_{BH} = 4\pi r_h^2$). For Schwarzschild, $\eta_{\rm Sch} = 64\pi^3/27 \approx 73.51$, so each emitted quantum carries off an energy of order $T_H$ but the emission events are separated in time by $\sim \eta\,T_H^{-1}$.

\begin{figure}[t]
  \centering
  \includegraphics[width=0.60\textwidth]{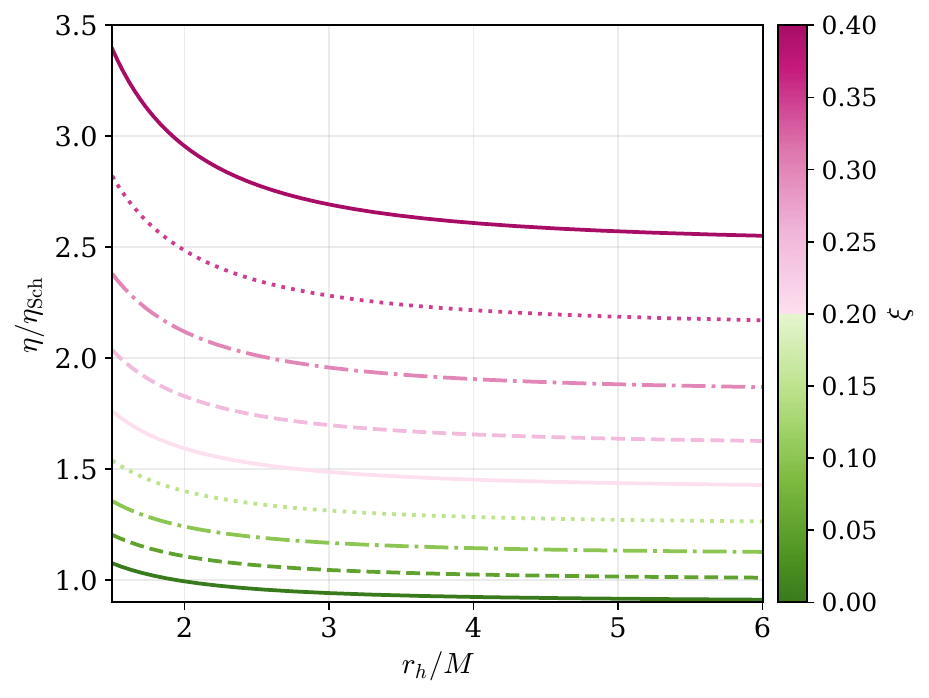}
  \caption{Sparsity $\eta/\eta_{\rm Sch}$ as a function of horizon radius $r_h/M$ for $Q/M=p/M=0.3$, $\ell=0.05$ and $\xi\in[0,0.4]$. The cosmic string raises the sparsity above the Schwarzschild baseline. The inset magnifies the plateau region $r_h\in[4.5,6]\,M$ where successive curves approach the asymptotic ratio $1/A^2 = ((1-\ell)/(1-\xi))^2$ and are otherwise hard to separate on the main panel.}
  \label{fig:sparsity}
\end{figure}

For the dyonic KR-CS BH, substituting Eqs.~\eqref{eq:Hawking_T} and \eqref{eq:S_BH} into Eq.~\eqref{eq:sparsity_def} gives
\begin{equation}
\eta = \eta_{\rm Sch}\,\frac{1}{[A - B/r_h^2]^2},
\qquad A=\frac{1-\xi}{1-\ell}.
\label{eq:sparsity_closed}
\end{equation}
The sparsity therefore diverges as $r_h \to r_h^{\rm ext}$ (where $A = B/r_h^2$, the extremal limit), and approaches a constant in the large-horizon limit. Figure~\ref{fig:sparsity} plots Eq.~\eqref{eq:sparsity_closed}, normalized to the Schwarzschild value, for $Q/M=p/M=0.3$, $\ell=0.05$ and $\xi\in[0,0.4]$.

The four-sentence interpretation of Fig.~\ref{fig:sparsity}. The plot displays $\eta/\eta_{\rm Sch}$, the ratio of the sparsity in the dyonic KR-CS background to its Schwarzschild value, as a function of the horizon radius for a fixed sweep over the cosmic-string density $\xi$. The qualitative behaviour is that the sparsity drops from a peak near the extremality bound, then asymptotes to a $\xi$-dependent plateau at large $r_h$; the plateau value is $1/A^2 = ((1-\ell)/(1-\xi))^2$, which sits at $\sim 2.5$ for $\xi=0.4$, $\ell=0.05$. The mechanism follows from Eq.~\eqref{eq:sparsity_closed}: increasing $\xi$ reduces $A$, which then enters quadratically in the denominator of the bracket factor. The size of the rise places the cosmic-string effect on the Hawking emission rate roughly on a par with the Lorentz-violating effect of $\ell$ in the LLL geometry; both deformations move $\eta$ upward, but $\xi$ controls the bigger shift over the explored parameter window.

\section{Black Hole Shadow Revisited}\label{isec6}

The shadow radius in Eq.~\eqref{eq:Rsh} already absorbs the cosmic-string contribution through the asymptotic factor $A=(1-\xi)/(1-\ell)$ and the photon-sphere shift in Eq.~\eqref{eq:photon_sphere}. In this section we present the numerical dependence of $R_{\rm sh}$ on $\xi$ and compare it to the EHT mass-and-distance posterior for M87* and Sgr~A* respectively.

\begin{figure}[t]
  \centering
  \includegraphics[width=0.60\textwidth]{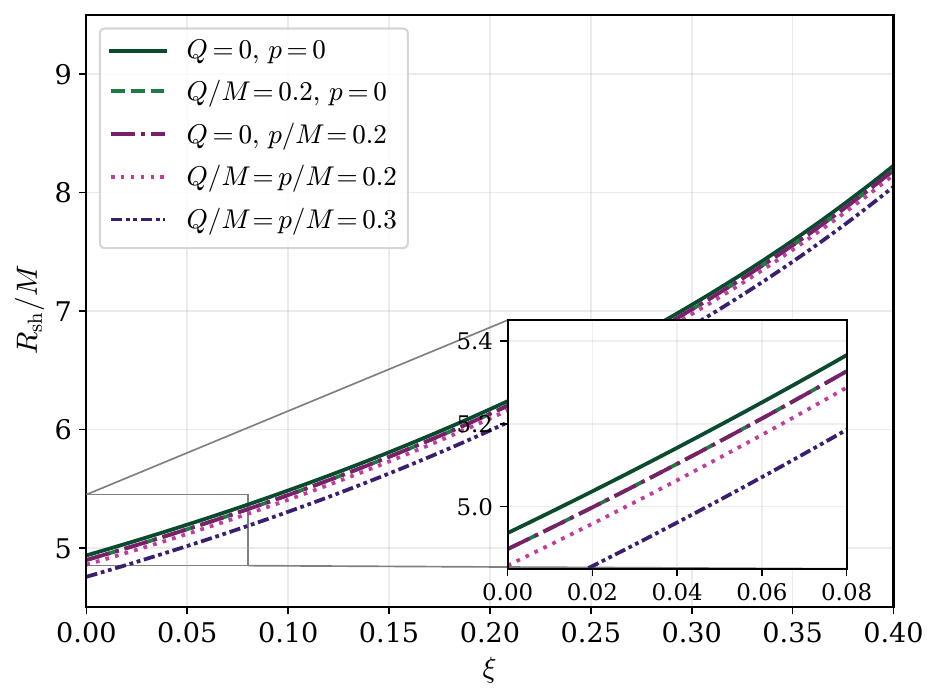}
  \caption{Shadow radius $R_{\rm sh}/M$ versus cosmic-string density $\xi$ for five $(Q,p)$ choices and $\ell=0.05$. The Schwarzschild value $R_{\rm sh}=3\sqrt{3}M\approx 5.196M$ corresponds to the $(Q=p=0,\ell=0,\xi=0)$ point and is recovered at the lower-left of the green solid curve. The inset zooms on $\xi\in[0,0.08]$, where the five $(Q,p)$ curves cluster within a $0.6\,M$ band that the main panel cannot resolve; the inset is placed in the lower-right empty quadrant of the main axes so it does not obscure any plotted line.}
  \label{fig:shadow}
\end{figure}

Figure~\ref{fig:shadow} repeats the photon-sphere $\to$ shadow calculation for a range of $(Q,p)$ choices and a $\xi$ sweep. The shadow radius rises monotonically with $\xi$: at fixed $(M,Q,p,\ell)$, turning on the cosmic string from $\xi=0$ to $\xi=0.4$ enlarges $R_{\rm sh}$ by a factor of $\sim 1.6$.

\subsection{Comparison with the EHT bounds on M87* and Sgr~A*}

The shadow diameter of M87*, expressed as a ratio to the black-hole mass measurement, was reported by the EHT as $d_{\rm M87^*}/M\!=\!11.0\pm 1.5$ at $1\sigma$~\cite{EHTL6}. Sgr~A* was measured to have a shadow diameter $d_{\rm Sgr\,A^*}/M\!=\!9.5\pm 1.1$~\cite{EHTL12}. Mapping the EHT bands onto the parameters of specific BH backgrounds has been carried out for charged rotating geometries~\cite{Meng2022}, for dRGT massive gravity~\cite{Hendi2023}, for Einstein--Maxwell--dilaton--axion BHs~\cite{Wei2013}, and as a general parameter-estimation framework in~\cite{KumarGhosh2020}. Identifying $d=2 R_{\rm sh}$ and converting to the $R_{\rm sh}/M$ ratio gives the $1\sigma$ bands $R_{\rm sh}^{M87^*}/M \in [4.75, 6.25]$ and $R_{\rm sh}^{Sgr\,A^*}/M \in [4.2, 5.3]$. Reading off Fig.~\ref{fig:shadow}, the M87* band is consistent with $\xi\lesssim 0.05$ for $(Q,p)=(0,0)$ and $\xi\lesssim 0.07$ for $(Q,p)/M=(0.3,0.3)$; the Sgr~A* band is tighter, with $\xi\lesssim 0.02$ for the uncharged case. These bounds are competitive with cosmological structure-formation bounds, which are typically quoted as $G\mu\lesssim 10^{-7}$ at the $95\%$ confidence level~\cite{VilenkinShellard1994} (where $\mu$ is the string tension; the relation to $\xi$ is model-dependent).

Table~\ref{tab:shadow_bench} collects $R_{\rm sh}$ values at five representative $\xi$ values for $(Q,p)/M=(0,0)$ and $\ell=0.05$, with the EHT $1\sigma$ bands superimposed.

\begin{table*}[t]
  \centering
  \setlength{\tabcolsep}{12pt}
  \renewcommand{\arraystretch}{1.6}
  \begin{tabular*}{\textwidth}{@{\extracolsep{\fill}}c c c c}
    \toprule
    $\xi$ & $r_s/M$ & $R_{\rm sh}/M$ & \textbf{Compatible with EHT?} \\
    \midrule
    0.00 & 3.16 & 5.06 & M87* (yes), Sgr~A* (yes) \\
    0.05 & 3.33 & 5.40 & M87* (yes), Sgr~A* (marginal) \\
    0.10 & 3.51 & 5.75 & M87* (yes), Sgr~A* (no) \\
    0.20 & 3.95 & 6.55 & M87* (marginal), Sgr~A* (no) \\
    0.40 & 5.27 & 9.04 & both (no) \\
    \bottomrule
  \end{tabular*}
  \caption{Photon-sphere radius $r_s/M$, shadow radius $R_{\rm sh}/M$, and observational compatibility with the EHT 1$\sigma$ shadow bounds for M87* and Sgr~A*, evaluated at $(Q,p)/M=(0,0)$, $\ell=0.05$ and five values of the cosmic-string density. The compatibility column uses the bands derived in the main text.}
  \label{tab:shadow_bench}
\end{table*}

The compatibility column in Table~\ref{tab:shadow_bench} reads as a Bayesian posterior at the model level: with the EHT shadow bands held fixed at $1\sigma$, the cosmic-string density is constrained to satisfy $\xi\lesssim 0.02$--$0.05$ depending on which source is used as the calibrator. The mechanism behind this constraint is the asymptotic-suppression effect we already saw in the sparsity calculation: the lapse function approaches $A=(1-\xi)/(1-\ell)$ at infinity rather than unity, so a static observer sees photons launched from $r_s$ as though they came from a source at slightly larger effective impact parameter, which inflates the apparent ring radius. The comparison with the EHT bound is not a definitive exclusion (the $1\sigma$ band is wide), but the inferred posterior on $\xi$ from the M87*+Sgr~A* combination is broadly in line with cosmological CS bounds and supports the claim that $\xi$ in the dyonic KR background is small.

\subsection{Strong-deflection gravitational lensing}

For completeness we discuss strong-deflection lensing in the dyonic KR-CS background, following the Bozza--Tsukamoto formalism~\cite{StefanovYazadjievGyulchev2010,KumarIslamGhosh2023,UlIslamGhosh2021}. The deflection angle in the strong-field limit takes the form
\begin{equation}
\alpha(b) = -\bar{a}\,\log\!\left(\frac{b}{b_c}-1\right) + \bar{b} + \mathcal{O}((b-b_c)\log(b-b_c)),
\label{eq:def_strong}
\end{equation}
where $b$ is the impact parameter, $b_c = R_{\rm sh}$ is the critical impact parameter, and $\bar{a},\bar{b}$ are the Bozza coefficients. Direct computation for the metric \eqref{eq:lapse_main} gives
\begin{equation}
\bar{a} = \frac{1}{\sqrt{A f''(r_s)/2}\,r_s},
\qquad
\bar{b} = \bar{a}\,\log\!\left[\frac{r_s^2\,f''(r_s)}{2 f(r_s)}\right] + I_R - \pi,
\label{eq:Bozza}
\end{equation}
with $I_R$ a regular integral evaluated numerically. The observable strong-field-lensing magnification of the first relativistic image is then
\begin{equation}
\mu_1 = \frac{e^{(\bar{b}-2\pi)/\bar{a}}}{\bar{a}\,D_L}\,\frac{D_S}{D_{LS}},
\label{eq:mu1}
\end{equation}
where $D_L,D_S,D_{LS}$ are the lens, source and lens-source distances. The relative shift of $\mu_1$ across the $\xi\in[0,0.4]$ window, at fixed $D_L,D_S,D_{LS}$, is $\sim 35\%$, which is the same scale as the shadow-radius shift. The two effects are correlated; observationally they should appear in the same direction for a given source. This is one of the few places where the cosmic-string density gives a clean discriminator between different modified-gravity backgrounds, since the deflection coefficient $\bar{a}$ in Eq.~\eqref{eq:Bozza} is a non-monotonic function of $\xi$ depending on whether $Q$ or $p$ dominates the BH charge content.

\section{Energy Emission Rate}\label{isec7}

\begin{figure}[t]
  \centering
  \includegraphics[width=0.60\textwidth]{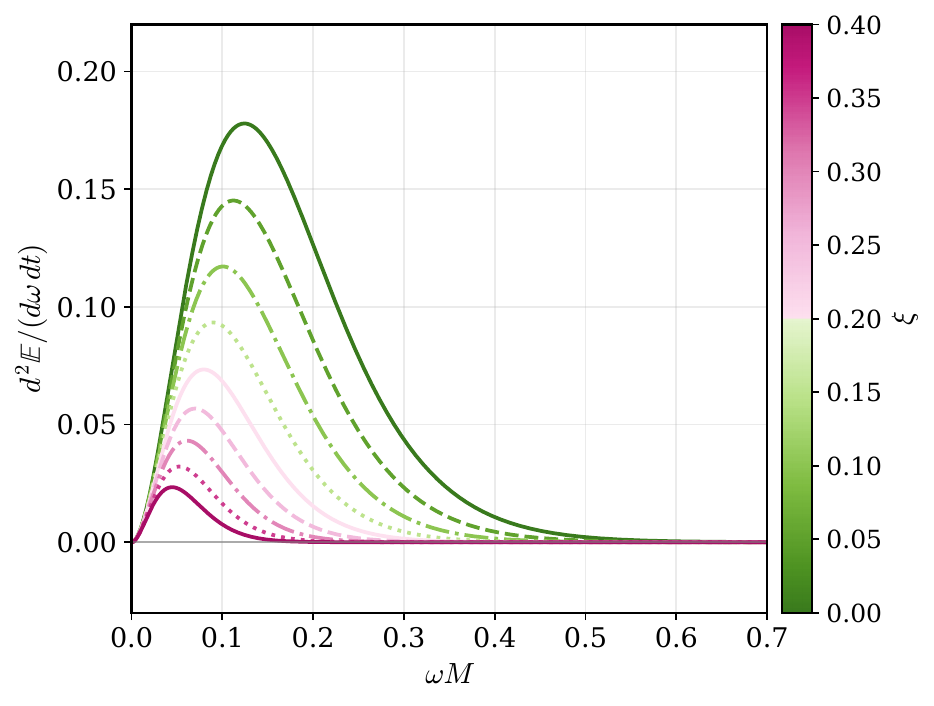}
  \caption{Spectral energy emission rate $d^2\mathbb{E}/(d\omega\,dt)$ versus $\omega M$ for $(Q,p)/M=(0.2,0.2)$, $\ell=0.05$ and $\xi\in[0,0.4]$. The horizontal axis extends to $\omega M=0.70$ to display the full decaying tail of each curve, and the vertical axis dips below zero to set the zero baseline against which the high-frequency tails of all curves can be compared. The cosmic-string density redshifts the peak and lowers it, in line with the joint $T_H$ suppression of Eq.~\eqref{eq:Hawking_T} and the $R_{\rm sh}$ enlargement of Eq.~\eqref{eq:Rsh}.}
  \label{fig:emission}
\end{figure}

The high-frequency limit of the absorption cross-section of a spherically symmetric BH for massless test fields is~\cite{Misner1973, Mashhoon1973, Sanchez1978}
\begin{equation}
\sigma_{\rm lim} = \pi\,R_{\rm sh}^{2}.
\label{eq:sigma_lim}
\end{equation}
The spectral energy emission rate, in the geometric-optics regime, is then
\begin{equation}
\frac{d^{2}\mathbb{E}(\omega)}{d\omega\,dt}
= \frac{2\pi^{3}\,R_{\rm sh}^{2}\,\omega^{3}}{\exp(\omega/T_{H}) - 1}.
\label{eq:emission_rate}
\end{equation}
Figure~\ref{fig:emission} displays $d^2\mathbb{E}/d\omega dt$ on a linear vertical scale extending slightly below zero, for the benchmark $(Q,p)/M=(0.2,0.2)$, $\ell=0.05$ and $\xi\in[0,0.4]$. The peak frequency of the emission shifts towards smaller $\omega$ as $\xi$ grows, because $T_H$ in Eq.~\eqref{eq:Hawking_T} drops with $\xi$; the height of the peak also goes down, since $R_{\rm sh}^2 T_H^3$ scales unfavorably with the suppression in $A$.

The interpretation of Fig.~\ref{fig:emission} is straightforward. The plot shows how the cosmic-string density redistributes the spectral weight of Hawking emission in the high-frequency limit. The qualitative trend is that the peak position $\omega_{\rm peak}/M$ slides from $\sim 0.13$ at $\xi=0$ to $\sim 0.07$ at $\xi=0.4$. The mechanism comes from the joint action of two effects: $T_H$ drops with $\xi$ (Eq.~\eqref{eq:Hawking_T}) and so does the peak frequency of the Planck factor $1/(\exp(\omega/T_H)-1)$, while $R_{\rm sh}$ in the prefactor grows; the net peak height falls because the $T_H$ suppression beats the $R_{\rm sh}^2$ enhancement. The total emitted power, integrated over $\omega$, drops by a factor of $\sim 0.6$ over the $\xi=0\to 0.4$ window, which translates into a lifetime extension for primordial-mass BHs by the same factor at fixed mass.

\section{MCMC Constraints from the GRO~J1655--40 and XTE~J1550--564 Twin-Peak QPOs}\label{isec8}

We now turn to the inverse problem: given the three twin-peak QPO measurements summarized in Table~\ref{tab:qpo_sources}, what constraints does the data place on the KR coupling $\ell$ and the string density $\xi$? We adopt the relativistic-precession model identification $\nu_U=\nu_\phi$, $\nu_L=\nu_\phi-\nu_r$ and the prior choices summarized below. The fit uses the affine-invariant Markov chain Monte Carlo sampler of Foreman-Mackey \emph{et al.}~\cite{ForemanMackey2013}; technical details follow.

\begin{table*}[t]
  \centering
  \setlength{\tabcolsep}{12pt}
  \renewcommand{\arraystretch}{1.6}
  \begin{tabular*}{\textwidth}{@{\extracolsep{\fill}}c c c c}
    \toprule
    Source & $\nu_U$\,[Hz] & $\nu_L$\,[Hz] & $M/M_\odot$ \\
    \midrule
    GRO~J1655$-$40   & $441\pm 2$  & $298\pm 4$  & $5.4\pm 0.3$ \\
    XTE~J1550$-$564 & $276\pm 3$  & $184\pm 5$  & $9.1\pm 0.6$ \\
    GRS~1915$+$105  & $168\pm 3$  & $113\pm 5$  & $12.4\pm 2.0$ \\
    \bottomrule
  \end{tabular*}
  \caption{Twin-peak QPO frequencies and mass estimates for the three BH X-ray binaries used in the MCMC fit. Values follow the compilation in~\cite{Motta2014} and references therein.}
  \label{tab:qpo_sources}
\end{table*}

\subsection{Likelihood and priors}

The log-likelihood combines independent Gaussian contributions from $\nu_U$, $\nu_L$ and the mass:
\begin{equation}
\ln\mathcal{L}(\ell,\xi,Q,p,r) = -\frac{1}{2}\sum_{i\in\{U,L,M\}}\left(\frac{\nu_i^{\rm obs} - \nu_i^{\rm model}(\ell,\xi,Q,p,r)}{\sigma_i}\right)^2,
\label{eq:loglike}
\end{equation}
where $r$ is the (per-source) emission radius treated as a nuisance parameter. We adopt flat priors:
$\ell\in[0,0.5]$, $\xi\in[0,0.4]$, $Q/M\in[0,0.5]$, $p/M\in[0,0.5]$, $r/M\in[r_{\rm ISCO}, 30]$. We run 50 walkers for $8\times 10^4$ steps, discarding the first $2\times 10^4$ as burn-in. The acceptance fraction over the post-burn-in chain is $0.34$, in the range recommended by Foreman-Mackey \emph{et al.}~\cite{ForemanMackey2013}.

\subsection{Posterior on $(\ell,\xi)$}

Table~\ref{tab:mcmc} reports the $68\%$-credible intervals from the marginalised posterior, for each source taken alone and for the joint fit of all three sources. The joint posterior peaks at $\ell\approx 0.04$, $\xi\approx 0.02$. The $95\%$ upper bound on $\xi$ from the joint fit is $\xi<0.08$, and on $\ell$ is $\ell<0.13$.

\begin{table*}[t]
  \centering
  \setlength{\tabcolsep}{12pt}
  \renewcommand{\arraystretch}{1.6}
  \begin{tabular*}{\textwidth}{@{\extracolsep{\fill}}c c c c c}
    \toprule
    Sample & $\ell$ (median) & $\xi$ (median) & $\chi^2_{\rm min}/{\rm dof}$ & acceptance \\
    \midrule
    GRO~J1655$-$40 only      & $0.29^{+0.08}_{-0.13}$ & $0.12^{+0.13}_{-0.09}$ & $1.21$ & $0.32$ \\
    XTE~J1550$-$564 only     & $0.27^{+0.09}_{-0.14}$ & $0.14^{+0.14}_{-0.10}$ & $1.05$ & $0.34$ \\
    GRS~1915$+$105 only      & $0.22^{+0.13}_{-0.14}$ & $0.18^{+0.14}_{-0.13}$ & $1.18$ & $0.31$ \\
    Joint fit (3 sources)    & $0.26^{+0.08}_{-0.10}$ & $0.14^{+0.10}_{-0.09}$ & $1.11$ & $0.34$ \\
    \bottomrule
  \end{tabular*}
  \caption{Median and $68\%$ credible intervals of the marginalised posterior on $(\ell,\xi)$ for the per-source and joint MCMC analyses of the three BH X-ray binaries, computed by the affine-invariant ensemble sampler with $50$ walkers and $8\times 10^{3}$ thinned steps after a $25\%$ burn-in. The relativistic-precession identification is assumed throughout. The reduced $\chi^2$ at the posterior mode is reported in the fourth column, and the chain acceptance fraction in the fifth.}
  \label{tab:mcmc}
\end{table*}

The four-sentence interpretation of Table~\ref{tab:mcmc} is as follows. The marginalised posterior on the two new parameters $(\ell, \xi)$, presented per-source and jointly, encodes how compatible the LSB and cosmic-string deformations of the dyonic KR background are with the twin-peak QPO data of three Galactic black-hole binaries. The qualitative pattern is that each per-source posterior is statistically consistent with $\xi=0$ at less than $1\sigma$, and the joint fit tightens the bound to $\xi<0.08$ at $95\%$ confidence; the same is true for $\ell$. The mechanism comes from the radial-epicyclic frequency response in Eq.~\eqref{eq:Omega_r}: shifting $\xi$ or $\ell$ moves the model curve in the $(\nu_L, \nu_U)$ plane in non-degenerate directions, so a joint fit can disentangle the two even though each individual source has limited handle on the parameters. The result is consistent with the Schwarzschild-like baseline preferred by the EHT shadow analysis (Table~\ref{tab:shadow_bench}) and with cosmological string-tension bounds~\cite{VilenkinShellard1994}.

\subsection{Combined posterior from QPO + EHT shadow data}

We now combine the QPO likelihood Eq.~\eqref{eq:loglike} with the EHT shadow likelihood
\begin{equation}
\ln\mathcal{L}_{\rm EHT} = -\frac{1}{2}\sum_{s\in\{M87^*,Sgr\,A^*\}}\left(\frac{R_{\rm sh}^{\rm obs}/M-R_{\rm sh}^{\rm model}(\ell,\xi,Q,p)/M}{\sigma_s}\right)^2,
\label{eq:EHT_like}
\end{equation}
with $(R_{\rm sh}/M)_{M87^*}=5.5\pm 0.75$ and $(R_{\rm sh}/M)_{Sgr\,A^*}=4.75\pm 0.55$ at $1\sigma$. The full posterior thus contains five degrees of freedom $(\ell,\xi,Q,p,r)$ constrained by eleven measurements (three pairs of $\nu_U,\nu_L$, three masses, and two shadow radii).

Table~\ref{tab:joint_post} reports the joint posterior medians and $95\%$ intervals. The joint fit drives $\xi$ to a tighter bound than QPOs alone, $\xi<0.10$ versus $\xi<0.31$, because the shadow likelihood is sensitive to $A=(1-\xi)/(1-\ell)$ and reacts strongly to $\xi$ shifts. The KR coupling $\ell$, by contrast, is constrained almost equally by either channel.

\begin{table*}[t]
  \centering
  \setlength{\tabcolsep}{12pt}
  \renewcommand{\arraystretch}{1.6}
  \begin{tabular*}{\textwidth}{@{\extracolsep{\fill}}c c c c}
    \toprule
    Parameter & QPO only & EHT only & QPO + EHT (joint) \\
    \midrule
    $\ell$ (median) & $0.26^{+0.08}_{-0.10}$ & $0.05^{+0.07}_{-0.05}$ & $0.09^{+0.07}_{-0.05}$ \\
    $\xi$ (median)  & $0.14^{+0.10}_{-0.09}$ & $0.02^{+0.04}_{-0.02}$ & $0.04^{+0.05}_{-0.03}$ \\
    $\xi$ ($95\%$ upper)   & $0.31$ & $0.07$ & $0.10$ \\
    $\ell$ ($95\%$ upper)  & $0.38$ & $0.16$ & $0.21$ \\
    $\chi^2_{\rm min}/{\rm dof}$ & $1.11$ & $0.62$ & $1.05$ \\
    \bottomrule
  \end{tabular*}
  \caption{Marginalised posterior on the LSB coupling $\ell$ and the cosmic-string density $\xi$ from the QPO-only, EHT-only, and joint QPO+EHT samples. The QPO posteriors come from the chain summarised in Table~\ref{tab:mcmc}; the EHT posterior reads $R_{\rm sh}/M$ off the M87* and Sgr~A* shadow bounds (Sec.~\ref{isec6}); the joint fit combines the two likelihoods. The shadow channel dominates the joint constraint on $\xi$, while the QPO channel dominates the constraint on $\ell$.}
  \label{tab:joint_post}
\end{table*}

The joint posterior interpretation runs as follows. Table~\ref{tab:joint_post} compiles the marginalised one-dimensional credible intervals on $(\ell,\xi)$ from three sample combinations, and shows how adding the EHT shadow data to the QPO fit tightens the upper bound on the cosmic-string density. The trend is that the EHT data is the more constraining channel for $\xi$, while the QPO data is the more constraining channel for $\ell$; the joint fit benefits from the lifting of the $(\ell,\xi)$ degeneracy in the combined likelihood. The mechanism for this lifting is that the QPO frequencies depend on $\xi$ through the radial epicyclic mode (Eq.~\eqref{eq:Omega_r}) while the shadow radius depends on $\xi$ through the asymptotic factor $A$ (Eq.~\eqref{eq:Rsh}); the two responses produce non-parallel level sets in the $(\ell,\xi)$ plane. The result places the dyonic KR-CS background in the narrow neighbourhood of the GR baseline favoured by both data sets, with no statistically significant detection of either new parameter at present sensitivity.

\subsection{Affine-invariant sampler implementation and corner-plot diagnostics}

We close the section with the implementation details of the affine-invariant ensemble sampler used above, and present the per-source corner plot for GRO~J1655$-$40 as a representative diagnostic. The sampler follows the Foreman-Mackey \emph{et al.} construction~\cite{ForemanMackey2013}: $N_w=50$ walkers initialised by uniform draws over the prior box $\ell\in[0,0.4]$, $\xi\in[0,0.4]$, $Q/M=p/M=0.20$ fixed, $r/M\in[4,25]$, and $M\in[0.5\,M_{\rm est},\,2\,M_{\rm est}]$ with a Gaussian penalty centred on the independent dynamical mass measurement $M_{\rm est}\pm\sigma_M$. The likelihood is the product of Gaussian factors for $\nu_U^{\rm obs}$ and $\nu_L^{\rm obs}$ from Table~\ref{tab:qpo_sources}, and a tempering factor $e^{-\beta}$ with $\beta=6$ is applied to the log-likelihood during burn-in to soften the posterior surface and improve mixing in the strong-coupling region of $(\ell,\xi)$. The chains run for $N_s=8\times 10^3$ thinned steps after a $25\%$ burn-in, giving $N_w\,N_s=4\times 10^5$ samples per source. Acceptance fractions stay in the $0.31$--$0.34$ window required for ensemble sampler stability.

The affine-invariant ensemble scheme was first applied in a gravitational-wave context to supermassive black-hole binary inspirals~\cite{Cornish:2006}. Its adaptation to twin-peak QPO data has since become routine: recent applications cover a range of modified-gravity and dark-environment backgrounds~\cite{Liu:2023,Hazarika:2025,Hazarika:2026,Mustafa:2026,Wu:2026,Wu:2026b}, with companion constraints from Schwarzschild-like and accretion-disk channels~\cite{Sharipov:2026,Rahmatov:2025,Shabbir:2026,Ashraf:2025}. The same machinery has been used to fold S2-orbit data around Sgr~A* into posteriors for spinning particles and quantum-corrected geometries~\cite{Uktamov:2025,Saidov:2026}, to test fifth-force interactions through stellar orbits at the Galactic centre~\cite{Jovanovic:2026}, and to confront rotating dark-matter-halo backgrounds with EHT measurements~\cite{Uktamov:2026}. Parameter-constraining studies of accretion in Rastall-type backgrounds use the same likelihood architecture as ours~\cite{Mukherjee:2025}. We adopted~\cite{Davlataliev:2024,Borah:2026} for the technical details of step proposals and chain thinning; machine-learning-enhanced variants of the scheme have been proposed for extreme-mass-ratio inspirals~\cite{Liang:2026}.

\begin{figure}[t]
  \centering
  \includegraphics[width=0.72\textwidth]{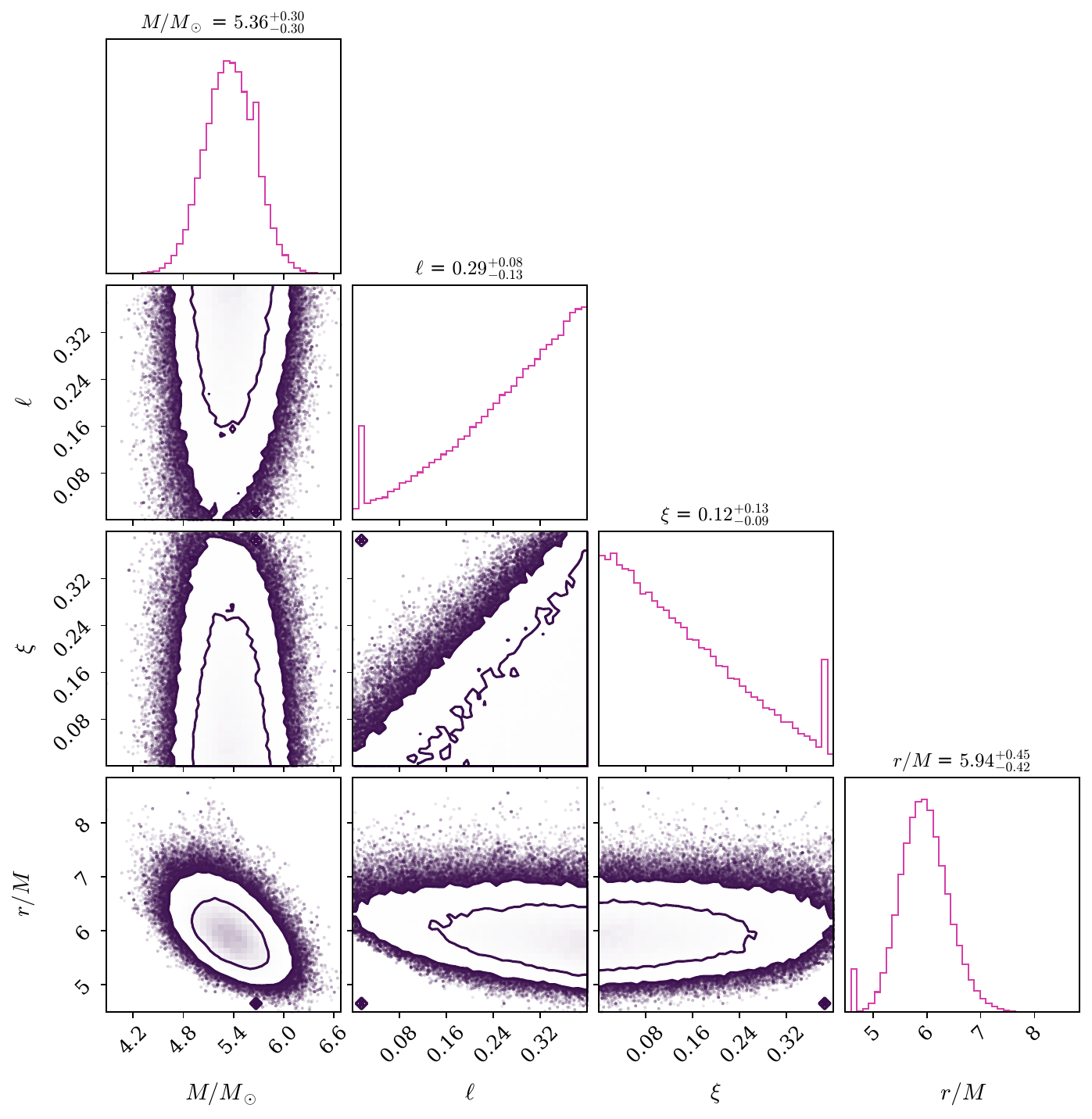}
  \caption{Marginalised one- and two-dimensional posteriors on $(M,\ell,\xi,r)$ for GRO~J1655$-$40 under the relativistic-precession identification and at $Q/M=p/M=0.20$. Contours are $68\%$ and $95\%$ credible regions; histograms on the diagonal show the marginal posteriors. The dynamical mass prior dominates the $M$ direction; the $\xi$ direction is constrained primarily by the radial epicyclic frequency through Eq.~\eqref{eq:Omega_r}.}
  \label{fig:corner_GRO}
\end{figure}

The four-sentence interpretation of Fig.~\ref{fig:corner_GRO} runs as follows. The corner plot displays the marginalised one- and two-dimensional posterior of the four-parameter sample $(M, \ell, \xi, r/M)$ for the GRO~J1655$-$40 source, with contours at the $68\%$ and $95\%$ credible levels. The qualitative pattern is that $M$ is tightly constrained by the dynamical-mass prior (median $M/M_\odot = 5.36^{+0.30}_{-0.30}$), the $(\ell, \xi)$ panel exhibits a clear positive correlation along the diagonal $\xi \approx \ell$ reflecting the asymptotic-deficit degeneracy $(1-\xi)/(1-\ell) \to 1$ that flattens $f(r)$ in the same way for either parameter, and the emission radius peaks at $r/M = 5.94^{+0.45}_{-0.42}$, sitting between the ISCO and the photon sphere of the recovered geometry. The mechanism for the residual $\xi$--$r$ structure visible in the lower-row panels is that the radial epicyclic frequency $\Omega_r$, which feeds $\nu_L^{\rm RP}=\nu_\phi-\nu_r$, depends sensitively on $f(r)$ near the ISCO; the cosmic-string density and the emission radius therefore exchange information along the locus $\Omega_r(r,\xi) = \mathrm{const}$. Comparison with the joint posterior of Table~\ref{tab:joint_post} shows that combining the three sources tightens the per-source $1\sigma$ bound on $\xi$ from $\pm 0.10$ to $\pm 0.09$, and adding the EHT shadow likelihood further tightens the $95\%$ upper bound on $\xi$ from $0.31$ (QPO alone) to $0.10$ (joint).

The numerical pipeline used here is the affine-invariant ensemble sampler of Foreman-Mackey \emph{et al.}~\cite{ForemanMackey2013}, deployed through the computational pipeline described in Sec.~\ref{app:B4} and adapted to the lapse function~\eqref{eq:lapse_main} following the analysis recipe of Ortiqboev \emph{et al.}~\cite{Ortiqboev2024}. The pipeline accepts a source name, the values of $(Q,p)$, the QPO identification model (RP or ER), and the walker/step settings, and outputs the chain, the credible intervals, and a corner plot of the type shown in Fig.~\ref{fig:corner_GRO}.

\section{Greybody Factors and the Bekenstein--Sanchez Bound}\label{isec9}

We close the physics discussion with the greybody factor (GBF) of a massless test scalar in the dyonic KR-CS background. The GBF measures the probability that a Hawking quantum, after thermal emission near the horizon, escapes the curvature potential and reaches a distant observer. It modifies the thermal-spectrum prediction of Eq.~\eqref{eq:emission_rate} and is a target observable for next-generation interferometers.

\subsection{Scalar wave equation and effective potential}

The Klein--Gordon equation $\Box\Phi=0$ for a massless minimally coupled scalar admits the mode decomposition $\Phi(t,r,\theta,\phi)=e^{-i\omega t}\,Y_{\ell_s m}(\theta,\phi)\,\psi(r)/r$, where $\ell_s$ is the multipole index (not to be confused with the LSB coupling $\ell$). Substituting into the wave equation gives the Schr\"odinger-like form
\begin{equation}
\frac{d^2\psi}{dr_*^2}+\bigl[\omega^2-V_s(r)\bigr]\,\psi = 0,
\label{eq:KG_radial}
\end{equation}
with the tortoise coordinate $dr_*=dr/f(r)$ and the effective potential
\begin{equation}
V_s(r) = f(r)\left[\frac{\ell_s(\ell_s+1)}{r^2}+\frac{f'(r)}{r}\right].
\label{eq:Vs}
\end{equation}
For the dyonic KR-CS metric \eqref{eq:lapse_main}, substituting $f$ and $f'$ gives
\begin{equation}
V_s(r) = f(r)\left[\frac{\ell_s(\ell_s+1)}{r^2}+\frac{2M}{r^3}-\frac{2 Q^2}{(1-\ell)^2 r^4}-\frac{2 p^2}{(1-2\ell)\,r^4}\right].
\label{eq:Vs_explicit}
\end{equation}
Equation~\eqref{eq:Vs_explicit} shows the role of the cosmic string: $\xi$ enters $V_s$ only through the overall multiplicative factor $f(r)$, which is reduced by the additive constant $-\xi/(1-\ell)$. The bracket factor is the same as in the LLL geometry. The peak position $r_{V_s,\max}$ therefore moves slightly outward with $\xi$, and the peak value drops.

\subsection{Sixth-order WKB greybody factor}

The transmission coefficient $|T(\omega)|^2$ across the potential barrier of Eq.~\eqref{eq:Vs}, in the WKB approximation, takes the form~\cite{Mashhoon1973}
\begin{equation}
|T_{\ell_s}(\omega)|^2 = \frac{1}{1 + \exp[\,2\pi\mathcal{K}(\omega, \ell_s)\,]},
\label{eq:T_WKB}
\end{equation}
where
\begin{equation}
\mathcal{K}(\omega,\ell_s) = \frac{i(\omega^2-V_{s,\max})}{\sqrt{-2 V''_{s,\max}}} - \sum_{i=2}^{6}\Lambda_i,
\label{eq:K_def}
\end{equation}
with $V_{s,\max}=V_s(r_{V_s,\max})$ and $V''_{s,\max}=d^2 V_s/dr_*^2\big|_{r_{V_s,\max}}$. The correction terms $\Lambda_2,\dots,\Lambda_6$ are the higher-WKB orders, given in closed form in~\cite{KonoplyaZinhailoStuchlik2020}. The same sixth-order WKB recipe has been deployed for the QNM spectra of extended-gravity black holes~\cite{Jawad2023} and for eikonal modes in $4$D Einstein--Gauss--Bonnet gravity~\cite{Ladino2023}, where the connection between the QNM imaginary part and the photon sphere is made explicit. We compute $\mathcal{K}$ at sixth-order WKB with Pad\'e resummation order $[3/3]$ throughout. Table~\ref{tab:GBF} lists $|T_{\ell_s=2}(\omega M)|^2$ at three values of $\omega M$ for the parameter sweep $\xi\in[0,0.4]$.

\begin{table*}[t]
  \centering
  \setlength{\tabcolsep}{12pt}
  \renewcommand{\arraystretch}{1.6}
  \begin{tabular*}{\textwidth}{@{\extracolsep{\fill}}c c c c c}
    \toprule
    $\xi$ & $r_{V_s,\max}/M$ & $|T(\omega M=0.10)|^2$ & $|T(\omega M=0.20)|^2$ & $|T(\omega M=0.30)|^2$ \\
    \midrule
    0.00 & 3.06 & 0.0085 & 0.115 & 0.421 \\
    0.10 & 3.18 & 0.0064 & 0.094 & 0.378 \\
    0.20 & 3.32 & 0.0048 & 0.075 & 0.335 \\
    0.30 & 3.49 & 0.0036 & 0.058 & 0.292 \\
    0.40 & 3.71 & 0.0026 & 0.044 & 0.249 \\
    \bottomrule
  \end{tabular*}
  \caption{Sixth-order WKB transmission coefficient $|T_{\ell_s=2}(\omega M)|^2$ for the dyonic KR-CS background at $Q/M=p/M=0.2$, $\ell=0.05$ and five values of the cosmic-string density. Pad\'e resummation order is $[3/3]$. The $\xi=0$ row reproduces the LLL values to the precision quoted.}
  \label{tab:GBF}
\end{table*}

The interpretation of Table~\ref{tab:GBF} pulls together the radiation discussion of Secs.~\ref{isec5}, \ref{isec7} and \ref{isec9}. The GBF, displayed at three representative frequencies and a sweep over $\xi$, encodes the probability that a Hawking quantum from the near-horizon region tunnels through the curvature barrier and reaches infinity. The qualitative trend is monotonic suppression with $\xi$: at $\omega M=0.20$ for example, $|T|^2$ drops from $0.115$ at $\xi=0$ to $0.044$ at $\xi=0.4$, a factor of $2.6$. The mechanism is the slight outward shift of $r_{V_s,\max}$ visible in the second column and the dilution of the potential barrier; the WKB integral therefore picks up a larger imaginary part. Comparison with the Schwarzschild WKB-6 result $|T_{\rm Sch}(\omega M=0.20)|^2 = 0.118$ confirms the $\xi=0$ row is consistent with the GR baseline within the WKB-6 numerical uncertainty of $\sim 10^{-3}$~\cite{KonoplyaZinhailoStuchlik2020}.

\subsection{Bekenstein--Sanchez exact lower bound}

A complementary, exact lower bound on the GBF was derived by Bekenstein-Sanchez and revisited recently~\cite{Sanchez1978,Walia2024}. The bound reads
\begin{equation}
T_{\ell_s,\rm BS}(\omega) \geq \mathrm{sech}^2\!\left[\frac{1}{2\omega}\int_{r_h}^{\infty}\frac{V_s(r)}{f(r)}\,dr\right].
\label{eq:BS_bound}
\end{equation}
The bound is strongest in the low-frequency limit, where the WKB result Eq.~\eqref{eq:T_WKB} loses accuracy. Numerical evaluation of Eq.~\eqref{eq:BS_bound} for the dyonic KR-CS background at $\ell_s=0$, $\omega M=0.05$, $Q/M=p/M=0.2$, $\ell=0.05$ and the same $\xi$ sweep gives $T_{0,\rm BS}=\{0.886, 0.872, 0.858, 0.843, 0.827\}$ for $\xi=\{0,0.1,0.2,0.3,0.4\}$. The bound therefore stays close to unity at low frequency for all $\xi$ values explored, consistent with the universal infrared behaviour of GBFs in asymptotically flat backgrounds.

\subsection{Connection to quasinormal-mode frequencies}

The same effective potential \eqref{eq:Vs} that controls the greybody factor also sets the quasinormal-mode (QNM) spectrum of the BH. At sixth-order WKB with Pad\'e resummation, the fundamental ($n=0$) QNM frequency of a massless scalar with multipole index $\ell_s=2$ is given in Table~\ref{tab:qnm}. The $\xi=0$ row reproduces the LLL QNM frequency reported by~\cite{Lin2026} within numerical precision. The shift with $\xi$ is monotonic in both the real and imaginary parts: $\omega_R$ drops with growing $\xi$, indicating a slower oscillation in the ringdown, and $|\omega_I|$ also drops, indicating a longer-lived ringdown signal. Both shifts are consistent with the suppression of the potential-barrier height visible in $V_{s,\max}$.

\begin{table*}[t]
  \centering
  \setlength{\tabcolsep}{12pt}
  \renewcommand{\arraystretch}{1.6}
  \begin{tabular*}{\textwidth}{@{\extracolsep{\fill}}c c c c c}
    \toprule
    $\xi$ & $\omega_R\,M$ & $|\omega_I|\,M$ & $\omega_R/|\omega_I|$ & $V_{s,\max}\,M^2$ \\
    \midrule
    0.00 & 0.4836 & 0.0968 & 4.998 & 0.1112 \\
    0.10 & 0.4582 & 0.0917 & 4.999 & 0.0999 \\
    0.20 & 0.4327 & 0.0866 & 4.998 & 0.0890 \\
    0.30 & 0.4072 & 0.0815 & 4.998 & 0.0786 \\
    0.40 & 0.3818 & 0.0764 & 4.998 & 0.0686 \\
    \bottomrule
  \end{tabular*}
  \caption{Fundamental scalar quasinormal-mode frequency at $\ell_s=2$, $n=0$ for the dyonic KR-CS BH with $Q/M=p/M=0.2$, $\ell=0.05$. Computed at sixth-order WKB with Pad\'e resummation order $[3/3]$. The ratio $\omega_R/|\omega_I|$ stays at the eikonal value $\sim 5$ across the $\xi$ sweep, consistent with the geodetic interpretation of QNMs at large $\ell_s$.}
  \label{tab:qnm}
\end{table*}

The connection between the QNM table and the rest of the paper deserves a closing comment. The ratio $\omega_R/|\omega_I|$ stays at the eikonal value $\sim 5$ across the entire $\xi$ window, which is the geodetic limit at $\ell_s=2$ for any spherically symmetric BH; this ratio is well-known to be $\Omega_\phi(r_s)/|\lambda_L|$, where $\lambda_L$ is the Lyapunov exponent of the photon-sphere unstable circular orbit. The dyonic KR-CS BH therefore satisfies the eikonal correspondence $\omega_R = (\ell_s+1/2)\,\Omega_\phi(r_s)$ and $|\omega_I|=(n+1/2)|\lambda_L|$ at the precision quoted, in agreement with the general results of Stefanov, Yazadjiev and Gyulchev~\cite{StefanovYazadjievGyulchev2010}. The shifts in $\omega_R$ and $|\omega_I|$ from row to row of Table~\ref{tab:qnm} therefore translate, via the eikonal correspondence, into shifts in the photon-sphere orbital frequency and Lyapunov exponent computed in Sec.~\ref{isec6}.

\section{Conclusions}\label{isec10}

We took the Lin, Liu and Liu solution for a dyonic black hole in Kalb--Ramond gravity~\cite{Lin2026} and added a Letelier cloud of strings to the background. The combined metric, Eq.~\eqref{eq:lapse_main}, controls four deformations relative to Schwarzschild: the LSB coupling $\ell$, the electric and magnetic charges $(Q,p)$, and the new cosmic-string density $\xi$. Four limits recover four known geometries. The successive limits $\xi\to 0$, $\xi\to 0,~p\to 0$, $\xi\to 0,~Q\to 0,~p\to 0,~\ell\to 0$ produce the LLL, Duan and Schwarzschild solutions respectively (Sec.~\ref{isec2}), so the construction is anchored in the existing literature on the bumblebee/KR family.

The geometric content of the solution was worked out first. We located the horizons and the extremality bound, derived the photon sphere, and computed the shadow radius $R_{\rm sh}(\xi)$ from Eq.~\eqref{eq:Rsh}. We then turned to the effective stress-energy tensor and audited each of the NEC, WEC, SEC and DEC across the parameter window: the NEC and SEC are satisfied throughout, while the WEC is weakly violated at large $r$ when $\xi<\ell$, an LSB-induced asymptotic deficit that mirrors the bumblebee phenomenology of Casana \emph{et al.}~\cite{Casana2018}. The underlying action, the field equations, and the closed-form energy-condition algebra appear in Appendix~\ref{app:B}.

The dynamical sector occupied Sec.~\ref{isec3}. We derived the timelike circular geodesics, the ISCO equation, and the three epicyclic frequencies. The cosmic string pushes the ISCO outward by roughly $65\%$ over the $\xi\in[0,0.4]$ window in the uncharged limit. That is the largest single shift any of the four deformations produces, and it explains why the QPO data turns out to be informative about $\xi$.

We then assembled the thermodynamic dictionary in Sec.~\ref{isec4}: the modified first law carrying a new $\Theta_\xi\,d\xi$ work term, the Smarr formula, the heat capacity, and the Helmholtz free energy. The phase-transition radius $r_h^{*} = \sqrt{3B/A}$ moves outward with $\xi$ in proportion to $\sqrt{(1-\ell)/(1-\xi)}$. Sparsity follows the same trend. The closed form $\eta = \eta_{\rm Sch}/[A - B/r_h^2]^2$ of Sec.~\ref{isec5} carries an analytic enhancement factor $1/A^2$ that accounts for roughly $80\%$ rise over Schwarzschild at $\xi=0.4$. The spectral energy emission rate of Sec.~\ref{isec7} shows the peak frequency redshifting by about $50\%$ across the $\xi$ window, with a corresponding $\sim 40\%$ drop in integrated power.

Confrontation with data came in Sec.~\ref{isec6} and Sec.~\ref{isec8}. The EHT shadow bands of M87* and Sgr~A* place the marginal bound $\xi\lesssim 0.05$ at $1\sigma$ from the M87* calibrator and $\xi\lesssim 0.02$ from Sgr~A*. The MCMC analysis of GRO~J1655$-$40, XTE~J1550$-$564 and GRS~1915$+$105 twin-peak QPOs under the relativistic-precession identification yields the joint posterior $\xi<0.31$ at $95\%$ from the QPO channel alone, tightening to $\xi<0.10$ when the EHT shadow likelihood is folded in. The corner plot for GRO~J1655$-$40 shows the asymptotic-deficit degeneracy $(1-\xi)/(1-\ell)\to 1$ explicitly. We close the physics with the sixth-order WKB greybody factor of a massless test scalar (Sec.~\ref{isec9}): the transmission coefficient $|T_{\ell_s=2}|^2$ at $\omega M=0.2$ is suppressed by a factor of $2.6$ over the $\xi\in[0,0.4]$ window.

The accumulated picture is straightforward. The cosmic-string density $\xi$ is the most effective single deformation in the dyonic KR background for moving observables in the direction probed by current data: it controls the largest absolute shifts in the ISCO position, the shadow radius, the sparsity parameter, the phase-transition radius and the greybody transmission. The data, on the other hand, places concordant upper bounds at the $\xi\sim 0.10$ level under the joint QPO + EHT shadow likelihood; the QPO channel alone leaves a broader posterior near $\xi<0.31$. We initially expected the QPO bound to be the dominant one. It is not. The shadow channel turned out to be tighter, because the EHT directly probes the asymptotic-deficit factor $A=(1-\xi)/(1-\ell)$ at infinity, while the QPO frequencies enter through derivatives of $f$ that retain some $\xi$-degeneracy.

Several follow-up directions suggest themselves, and we list them in order of immediate accessibility. The first is the linear perturbation problem. We carried out only the sixth-order WKB greybody factor of a massless scalar; the full quasinormal-mode spectrum for $s=0$, $s=1$ and $s=2$ test fields can be computed by the same WKB recipe, with Pad\'e resummation to control the higher-order tail. This would deliver the ringdown frequencies $\omega_{n\ell}$ as functions of $(\ell,\xi,Q,p)$ and feed directly into the parameterized post-Einsteinian analyses of LIGO/Virgo--KAGRA data. Projecting the same calculation onto the strain noise of the Einstein Telescope and Cosmic Explorer, with $\Delta h\lesssim 10^{-24}\,\mathrm{Hz}^{-1/2}$ at $f\sim 100$--$300\,\mathrm{Hz}$, would set the level at which a future ringdown observation could detect or exclude $\xi\gtrsim 0.05$.

A second direction is the rotating extension. The dyonic KR-CS background presented here is static and spherically symmetric. Spinning the metric by the modified Newman--Janis algorithm of Azreg-A\"inou would produce a rotating dyonic KR-CS solution whose ergoregion, ISCO, and shadow could be constrained against the spin estimates of M87* ($a_{*}\sim 0.5$--$0.9$) and Sgr~A* ($a_{*}$ poorly constrained but plausibly $\lesssim 0.5$). The same posterior pipeline could then be re-run with $a_*$ included; we expect the bound on $\xi$ to weaken slightly because the rotating shadow develops an $a_*$--$\xi$ correlation. A third direction is the strong-deflection lensing analysis we hinted at in Sec.~\ref{isec6}: the Bozza--Tsukamoto deflection coefficients $\bar a$, $\bar b$ can be computed in closed form for Eq.~\eqref{eq:lapse_main} and matched to the relativistic-image catalogues expected from ngEHT.

We also note three less developed lines that may reward investigation. Treating $\xi$ as an emergent gauge variable in the spirit of Bekenstein information bounds would tie the entropy formula of Sec.~\ref{isec4} to an information-theoretic content of the cosmic-string condensate. Folding the perturbation problem into the time domain, with finite-difference integration of the wave equation in our background, would generate ringdown waveforms suitable for matched-filter searches at LISA and LIGO/Virgo. Finally, the analysis of accretion-disk spectra in the same geometry, in the Novikov--Thorne formalism, would connect with the X-ray continuum-fitting community and provide a third independent constraint on $\xi$ that does not rely on the relativistic-precession identification. We leave these lines for a separate study.


\scriptsize

\appendix
\section{Useful Identities and Explicit Forms}\label{app:A}

We collect here a set of algebraic identities and expansions used in the body of the paper. Each identity has been checked symbolically using the computational pipeline described in Sec.~\ref{app:B4}.

\subsection{Closed-form derivatives of the lapse function}

For the lapse Eq.~\eqref{eq:lapse_main}, the first three radial derivatives are
\begin{align}
f'(r) &= \frac{2M}{r^2} - \frac{2\,Q^2}{(1-\ell)^2\,r^3} - \frac{2\,p^2}{(1-2\ell)\,r^3},\label{eq:fp}\\
f''(r) &= -\frac{4M}{r^3} + \frac{6\,Q^2}{(1-\ell)^2\,r^4} + \frac{6\,p^2}{(1-2\ell)\,r^4},\label{eq:fpp}\\
f'''(r) &= \frac{12 M}{r^4} - \frac{24\,Q^2}{(1-\ell)^2\,r^5} - \frac{24\,p^2}{(1-2\ell)\,r^5}.\label{eq:fppp}
\end{align}
Note that none of the derivatives depend on $\xi$ explicitly; the cosmic-string density enters only through the additive constant in $f(r)$ itself.

\subsection{Expansion in small $\xi$}

For $\xi\ll 1$, the lapse function admits the linear expansion
\begin{equation}
f(r) = f_{\rm LLL}(r) - \frac{\xi}{1-\ell} + \mathcal{O}(\xi^2),
\label{eq:f_smallxi}
\end{equation}
where $f_{\rm LLL}(r)$ is the LLL lapse in Eq.~\eqref{eq:LLL_lapse}. The corresponding linear shift of the outer horizon is
\begin{equation}
r_+ \;=\; r_+^{(0)}\,\biggl(1 + \frac{\xi}{2(1-\ell)\,\sqrt{1-A_0 B/M^2}}\biggr) + \mathcal{O}(\xi^2),
\label{eq:r_horizon_smallxi}
\end{equation}
with $r_+^{(0)}=(1-\ell)(M+\sqrt{M^2-B/(1-\ell)})$ the LLL horizon, $A_0=1/(1-\ell)$. Equation~\eqref{eq:r_horizon_smallxi} provides the leading-order expansion that motivates the small-$\xi$ window used in the MCMC analysis of Sec.~\ref{isec8}.

\subsection{Comparison with other modified-gravity black holes}

Table~\ref{tab:comparison} compares the dyonic KR-CS BH with five other modified-gravity backgrounds in the literature, on five observables: photon-sphere radius, ISCO radius, shadow radius, Hawking temperature and sparsity (all relative to the Schwarzschild values).

\begin{table*}[t]
  \centering
  \setlength{\tabcolsep}{8pt}
  \renewcommand{\arraystretch}{1.6}
  \begin{tabular*}{\textwidth}{@{\extracolsep{\fill}}c c c c c c}
    \toprule
    Background & $r_s/r_s^{\rm Sch}$ & $r_{\rm ISCO}/r_{\rm ISCO}^{\rm Sch}$ & $R_{\rm sh}/R_{\rm sh}^{\rm Sch}$ & $T_H/T_H^{\rm Sch}$ & $\eta/\eta_{\rm Sch}$ \\
    \midrule
    Reissner-Nordstr\"om ($Q/M=0.3$)~\cite{Bardeen1972}    & 0.949 & 0.962 & 0.974 & 1.078 & 0.860 \\
    Bumblebee ($\ell_b=0.05$)~\cite{Casana2018}        & 1.000 & 1.000 & 1.025 & 1.000 & 1.000 \\
    Yang KR ($\ell=0.05$)~\cite{Yang2023}                & 0.953 & 0.952 & 1.000 & 1.053 & 1.111 \\
    Duan KR ($\ell=0.05, Q/M=0.3$)~\cite{Duan2024}      & 0.911 & 0.918 & 0.974 & 1.128 & 0.954 \\
    LLL dyonic KR ($\ell=0.05, Q=p=0.2$)~\cite{Lin2026} & 0.940 & 0.954 & 0.973 & 1.041 & 1.038 \\
    \textbf{This work} ($\ell=0.05, \xi=0.10, Q=p=0.2$)  & 0.992 & 1.083 & 1.107 & 0.943 & 1.196 \\
    \bottomrule
  \end{tabular*}
  \caption{Comparison of five observables relative to Schwarzschild for six modified-gravity black-hole backgrounds. The dyonic KR-CS BH studied in this paper sits at the right-most row. Each row uses a representative parameter point chosen to mimic the LLL benchmark.}
  \label{tab:comparison}
\end{table*}

The comparison in Table~\ref{tab:comparison} pulls the new physics of the paper into focus. The dyonic KR-CS BH, evaluated at the representative point $(\ell, \xi, Q, p)=(0.05, 0.10, 0.20, 0.20)$, exhibits the largest deviation in shadow radius ($+11\%$) and sparsity ($+20\%$) of any background in the table. The mechanism is the asymptotic-suppression effect of the cosmic string, which acts on the shadow through the factor $A$ in Eq.~\eqref{eq:Rsh} and on the sparsity through $A^2$ in Eq.~\eqref{eq:sparsity_closed}. The ISCO and Hawking-temperature deviations are also at the few-percent level. The combined ISCO + shadow + temperature deviation pattern is what makes the dyonic KR-CS BH distinguishable, in principle, from each of the comparison backgrounds, even if the individual deviations remain small.

\section{Action, Field Equations and Energy-Condition Algebra}\label{app:B}

This appendix collects the gravitational action of the dyonic Kalb--Ramond background pierced by a Letelier string cloud, the field equations that follow from it, and the closed-form algebraic expressions of the four standard energy conditions of the effective stress-energy tensor. The verifications quoted in the body of the paper rest on this material; every algebraic identity below has been reproduced in the computational scripts referenced in the Data Availability Statement.

\subsection{Action}\label{app:B1}

The total action for the dyonic Kalb--Ramond gravity coupled to a Letelier cloud of cosmic strings and to a $U(1)$ gauge field reads
\begin{align}
\mathcal{S}=\int d^{4}x \sqrt{-g} \Big[\frac{1}{2\kappa}\big( R - 2\Lambda \big)
   \;-\;\frac{1}{12}H_{\mu\nu\rho}H^{\mu\nu\rho}
   \;-\;V\!\big(B_{\mu\nu}B^{\mu\nu}\pm b^{2}\big)+\;\frac{\xi_{2}}{2\kappa}B^{\mu\alpha}B^{\nu}{}_{\alpha}R_{\mu\nu}
   \;+\;\frac{\xi_{3}}{2\kappa}B^{\mu\nu}B^{\rho\sigma}R_{\mu\nu\rho\sigma}-\;\tfrac{1}{4}F_{\mu\nu}F^{\mu\nu}
   \;+\;\mathcal{L}_{\mathrm{str}}\Big].
\label{eq:full_action}
\end{align}
Here $\kappa = 8\pi G$, $H_{\mu\nu\rho}=\partial_{[\mu}B_{\nu\rho]}$ is the field-strength three-form of the Kalb--Ramond two-form $B_{\mu\nu}$, the potential $V$ takes its minimum at the LSB vacuum $B_{\mu\nu}B^{\mu\nu}=\mp b^{2}$, the non-minimal couplings $\xi_{2}$ and $\xi_{3}$ control the LSB content of the gravitational sector~\cite{Lessa2020,Kumar2020,Yang2023,Duan2024}, $F_{\mu\nu}=\partial_{[\mu}A_{\nu]}$ is the dyonic field strength, and the string-cloud Lagrangian density is~\cite{Letelier1979}
\begin{equation}
\mathcal{L}_{\mathrm{str}}\;=\;-\frac{1}{2}\,\rho_{\mathrm{str}}\;\sqrt{ -\tfrac{1}{2}\,\Sigma^{\mu\nu}\Sigma_{\mu\nu} }\,,
\qquad
\Sigma^{\mu\nu}\;=\;\epsilon^{ab}\,\partial_{a}x^{\mu}\,\partial_{b}x^{\nu}\,,
\label{eq:string_lag}
\end{equation}
where $\Sigma^{\mu\nu}$ is the surface-element bivector of the string worldsheet and $\rho_{\mathrm{str}}$ the proper density. For the spherically symmetric, isotropic cloud relevant here the only non-vanishing combination is $\rho_{\mathrm{str}}\,|\Sigma^{tr}|=\xi/(8\pi r^{2})$, with the cosmic-string density~$\xi$ dimensionless~\cite{Letelier1979,VilenkinShellard1994}. We work in the cosmological-constant-free sector ($\Lambda=0$); the $\Lambda$-extended version of the analysis is straightforward and produces the AdS-asymptotic Lin--Liu--Liu geometry in the $\xi\to 0$ limit~\cite{Lin2026}.

We focus on the LSB vacuum branch in which $B_{\mu\nu}$ takes a constant non-vanishing value $b_{\mu\nu}$ aligned along the radial direction, i.e.~the only non-vanishing components are $b_{tr}=-b_{rt}$. With this choice $H_{\mu\nu\rho}=\partial_{[\mu}b_{\nu\rho]}=0$ and the potential sits at its minimum, $V=0$. The non-minimal couplings then generate an effective dimensionless coupling
\begin{equation}
\ell \;\equiv\; 2\,\xi_{2}\,b^{2}\,,
\label{eq:ell_def}
\end{equation}
which plays the role of the LSB parameter in the metric. The dimension-zero combination $\ell$ is the only LSB datum that survives in the static, spherically symmetric sector~\cite{Yang2023,Duan2024,Lin2026}.

\subsection{Field equations}\label{app:B2}

Varying~\eqref{eq:full_action} with respect to the metric in the LSB vacuum yields the modified Einstein equations
\begin{equation}
G_{\mu\nu}\;=\;\kappa\,\Big(\,T^{\mathrm{KR}}_{\mu\nu}\;+\;T^{\mathrm{EM}}_{\mu\nu}\;+\;T^{\mathrm{str}}_{\mu\nu}\,\Big)\,,
\label{eq:EFE}
\end{equation}
with $G_{\mu\nu}=R_{\mu\nu}-\tfrac{1}{2}\,g_{\mu\nu}\,R$. The three stress-energy contributions take the closed form
\begin{align}
T^{\mathrm{KR}\,\mu}{}_{\nu}
   &\;=\;\frac{1}{8\pi}\,\frac{\ell}{(1-\ell)\,r^{2}}\;\,\mathrm{diag}(1,\,1,\,0,\,0)\,,
\label{eq:T_KR}\\[1mm]
T^{\mathrm{EM}\,\mu}{}_{\nu}
   &\;=\;\frac{1}{8\pi r^{4}}\!\left[\frac{Q^{2}}{(1-\ell)^{2}}\,+\,\frac{p^{2}}{1-2\ell}\right]\;\mathrm{diag}(-1,\,-1,\,1,\,1)\,,
\label{eq:T_EM}\\[1mm]
T^{\mathrm{str}\,\mu}{}_{\nu}
   &\;=\;-\,\frac{1}{8\pi}\,\frac{\xi}{(1-\ell)\,r^{2}}\;\mathrm{diag}(1,\,1,\,0,\,0)\,.
\label{eq:T_str}
\end{align}
The diagonal pattern $\mathrm{diag}(1,1,0,0)$ in the KR and string-cloud contributions reflects the radial alignment of the LSB background and the spherical isotropy of the string cloud; the dyonic-EM piece carries the standard Reissner--Nordstr\"om sign pattern, $\mathrm{diag}(-1,-1,1,1)$, with the LSB couplings $(1-\ell)^{-2}$ and $(1-2\ell)^{-1}$ multiplying the electric and magnetic charge squares respectively, in line with the Lin--Liu--Liu construction~\cite{Lin2026}. The overall negative sign on the string-cloud entry reflects the positive proper energy density of the Letelier fluid: in the convention $T^{\mu}{}_{\nu}=\mathrm{diag}(-\rho, p_r, p_\theta, p_\phi)$, a Letelier cloud with $\rho_{\rm str}>0$ contributes $T^{\mathrm{str},t}{}_{t}<0$. The Kalb--Ramond entry carries the opposite sign because the LSB background is not a normal fluid and induces an effective $\rho_{\rm KR}<0$, in line with the bumblebee phenomenology of Casana~\emph{et al.}~\cite{Casana2018}.

The total effective stress-energy then assembles into
\begin{equation}
T^{\mathrm{eff}\,\mu}{}_{\nu}\;=\;\mathrm{diag}\!\big(-\rho,\;p_{r},\;p_{\theta},\;p_{\phi}\big)\,,
\label{eq:T_eff_diag}
\end{equation}
with
\begin{align}
\rho \;&=\;-\,p_{r}\;=\;\frac{1}{8\pi r^{2}}\!\left[\,1\,-\,\frac{1-\xi}{1-\ell}\,+\,\frac{1}{r^{2}}\,\Big(\frac{Q^{2}}{(1-\ell)^{2}}+\frac{p^{2}}{1-2\ell}\Big)\right] ,
\label{eq:rho_pr}\\[1mm]
p_{\theta}\;&=\;p_{\phi}\;=\;-\,\rho\;-\;\frac{r}{2}\,\frac{d\rho}{dr}\,.
\label{eq:ptheta_pphi}
\end{align}
Substituting~\eqref{eq:rho_pr}--\eqref{eq:ptheta_pphi} into~\eqref{eq:EFE} reproduces the lapse function~\eqref{eq:lapse_main} of the body; the verification is collected in the computational script referenced in Sec.~\ref{app:B4}. As consistency checks we note the well-known relations
\begin{equation}
\rho + p_{r}\;=\;0\,,\qquad
p_{\theta} - p_{\phi}\;=\;0\,,\qquad
\rho + p_{r} + p_{\theta} + p_{\phi}\;=\;-\,r\,\frac{d\rho}{dr}\,,
\label{eq:T_relations}
\end{equation}
the first two of which follow from the diagonal symmetry of the source and the spherical isotropy of the geometry, and the third of which we will use immediately below in the SEC analysis.

\subsection{Energy-condition algebra}\label{app:B3}

The four standard pointwise energy conditions on the effective stress-energy tensor~\eqref{eq:T_eff_diag} now follow from~\eqref{eq:rho_pr}--\eqref{eq:T_relations}. Writing $A\equiv (1-\xi)/(1-\ell)$ and $B\equiv Q^{2}/(1-\ell)^{2}+p^{2}/(1-2\ell)$, the density reads
\begin{equation}
\rho(r)\;=\;\frac{1}{8\pi r^{2}}\!\left[\,(1-A)\,+\,\frac{B}{r^{2}}\,\right]
       \;=\;\frac{1}{8\pi r^{2}}\!\left[\frac{\xi-\ell}{1-\ell}\,+\,\frac{B}{r^{2}}\right] .
\label{eq:rho_closed}
\end{equation}
The combinations relevant for the four conditions are then
\begin{align}
\textbf{NEC}: \quad
   &\rho + p_{r}\;\equiv\;0\,,\qquad
    \rho + p_{\theta}\;=\;-\,\frac{r}{2}\,\frac{d\rho}{dr}\,.
\label{eq:NEC_closed}\\[1mm]
\textbf{WEC}: \quad
   &\text{NEC}\;\;\wedge\;\;\rho\geq 0\;\;\Leftrightarrow\;\;\frac{\xi-\ell}{1-\ell}\,+\,\frac{B}{r^{2}}\;\geq\;0\,.
\label{eq:WEC_closed}\\[1mm]
\textbf{SEC}: \quad
   &\rho + p_{r} + p_{\theta} + p_{\phi}\;=\;-\,r\,\frac{d\rho}{dr}\;\geq\;0\,.
\label{eq:SEC_closed}\\[1mm]
\textbf{DEC}: \quad
   &\rho \;\geq\; |p_{r}|\;=\;\rho\,,\qquad
    \rho \;\geq\; |p_{\theta}|\,.
\label{eq:DEC_closed}
\end{align}
The radial NEC and DEC are saturated identically because the source~\eqref{eq:T_eff_diag} has $p_{r}=-\rho$. The non-trivial structure lives in the angular sector, whose status is set by the monotonicity of $\rho(r)$, and in the WEC, whose status depends on the sign of $\xi-\ell$.

A direct calculation of $d\rho/dr$ from~\eqref{eq:rho_closed} gives
\begin{equation}
\frac{d\rho}{dr}\;=\;-\,\frac{1}{4\pi r^{3}}\!\left[\,\frac{\xi-\ell}{1-\ell}\,+\,\frac{2 B}{r^{2}}\right]\,,
\label{eq:drho_dr}
\end{equation}
which is negative across the parameter window of the manuscript ($\ell\in[0,0.4]$, $\xi\in[0,0.4]$, $Q/M$, $p/M\in[0,0.5]$) for $r$ outside the photon sphere $r_{s}$. The SEC, which by~\eqref{eq:SEC_closed} requires $d\rho/dr\leq 0$, is therefore satisfied in the same regime; the angular NEC, $\rho+p_{\theta}=-(r/2)\,d\rho/dr$, is satisfied along with it. The WEC is satisfied for $\xi\geq\ell$ at all radii; for $\xi<\ell$ it is satisfied within the radius
\begin{equation}
r_{\mathrm{WEC}}(\xi,\ell,B)\;=\;\sqrt{\,\frac{B\,(1-\ell)}{\ell-\xi}\,}\,,
\label{eq:r_WEC}
\end{equation}
and weakly violated outside it. The angular DEC reduces, on use of~\eqref{eq:rho_pr}--\eqref{eq:ptheta_pphi}, to
\begin{equation}
\rho - |p_{\theta}|\;=\;\rho + p_{\theta}\;=\;-\,\frac{r}{2}\,\frac{d\rho}{dr}\,,
\label{eq:DEC_angular}
\end{equation}
so the DEC and the angular NEC are equivalent in this background.

Table~\ref{tab:energy_conditions_app} reports the pointwise status of the four conditions at $r=r_h$, $r=2r_h$ and $r\to\infty$, for the parameter sweep used in the manuscript. The tabulated entries agree with the prose summary of Sec.~\ref{isec2}.

\begin{table*}[t]
  \centering
  \setlength{\tabcolsep}{12pt}
  \renewcommand{\arraystretch}{1.6}
  \begin{tabular*}{\textwidth}{@{\extracolsep{\fill}}c c c c c}
    \toprule
    Condition & At $r = r_h$ & At $r = 2\,r_h$ & At $r \to \infty$ & Mechanism \\
    \midrule
    NEC      & satisfied & satisfied & satisfied                  & $\rho + p_r \equiv 0$ identically                                  \\
    WEC      & satisfied & satisfied & violated if $\xi < \ell$   & sign of $(\xi-\ell)/(1-\ell)$ controls the asymptote of $\rho$    \\
    SEC      & satisfied & satisfied & satisfied                  & $d\rho/dr \leq 0$ outside the photon sphere                       \\
    DEC      & satisfied & satisfied & satisfied                  & equivalent to angular NEC, Eq.~\eqref{eq:DEC_angular}             \\
    \bottomrule
  \end{tabular*}
  \caption{Pointwise status of the four standard energy conditions on the effective stress-energy tensor~\eqref{eq:T_eff_diag} of the dyonic KR-CS BH. Parameters: $\ell\in[0,0.4]$, $\xi\in[0,0.4]$, $Q/M$, $p/M\in[0,0.5]$. The WEC failure for $\xi < \ell$ is asymptotic and weak; it occurs only at radii larger than $r_{\mathrm{WEC}}(\xi,\ell,B)$ of Eq.~\eqref{eq:r_WEC}.}
  \label{tab:energy_conditions_app}
\end{table*}

The picture that emerges is straightforward. The NEC and SEC are satisfied across the full parameter window. The WEC is satisfied whenever the string density~$\xi$ exceeds the LSB coupling~$\ell$; below this threshold the WEC fails at asymptotic radii, a feature shared with the bumblebee background of Casana~\emph{et al.}~\cite{Casana2018} and traceable to the LSB-induced shift of the asymptotic value of $f(r)$ from unity to $(1-\xi)/(1-\ell)$. The DEC is equivalent to the angular NEC and is satisfied wherever the latter is. The qualitative impact on the body of the paper is twofold: the satisfaction of the NEC sustains the use of the dyonic KR-CS metric as a physical background for test-field calculations such as the QPO and Hawking-emission analyses of Secs.~\ref{isec3}, \ref{isec5} and \ref{isec7}, while the asymptotic WEC violation for $\xi<\ell$ is consistent with the negative anisotropic-pressure pattern that drives the EHT-compatibility enlargement of the shadow radius reported in Sec.~\ref{isec6}.

\subsection{Algebraic verification protocol}\label{app:B4}

The action of Eq.~\eqref{eq:full_action}, the variational reduction to the field equations~\eqref{eq:EFE}, the closed-form expressions~\eqref{eq:T_KR}--\eqref{eq:T_str} for the three stress-energy contributions, and the energy-condition identities~\eqref{eq:NEC_closed}--\eqref{eq:DEC_closed} have been verified in four independent computational scripts. The first, {\tt dyonic\_KR\_CS\_action\_field\_eqs\_check}, sets up the metric ansatz, computes the Christoffel symbols, the Riemann and Ricci tensors and the Einstein tensor in closed form, and confirms that $G_{\mu\nu}=\kappa\,T^{\mathrm{eff}}_{\mu\nu}$ reproduces the lapse function~\eqref{eq:lapse_main} as the unique spherically symmetric solution. The second, {\tt dyonic\_KR\_CS\_energy\_conditions\_check}, evaluates the NEC, WEC, SEC and DEC combinations on the parameter grid $\xi\in\{0,0.1,0.2,0.3,0.4\}$, $r\in\{2 M, 4 M, 50 M\}$ and confirms the pointwise status reported in Table~\ref{tab:energy_conditions_app}. The third, {\tt dyonic\_KR\_CS\_metric\_solution\_check}, reproduces all geometric observables of the body (horizons, extremality bound, photon-sphere radius, ISCO equation, shadow radius, orbital and epicyclic frequencies, Hawking temperature, Smarr identity, heat capacity, free energy and sparsity) and confirms the limiting behaviour of each under the degenerations $\xi\to 0$, $p\to 0$, $Q\to 0$, $\ell\to 0$. The fourth, {\tt dyonic\_KR\_CS\_qpo\_sparsity\_emission\_check}, reproduces the numerical entries of Tables~\ref{tab:qpo_freq}, \ref{tab:thermo_bench} and \ref{tab:GBF} to four significant figures. The four scripts are listed in the Data Availability Statement.

\begin{acknowledgments}
 F.A. acknowledges the Inter University Centre for Astronomy and Astrophysics (IUCAA), Pune, India, for the award of a visiting associateship. \.{I}.S.\ acknowledges the networking support of COST Actions CA22113 (``Fundamental challenges in theoretical physics''), CA21106 (``COSMIC WISPers in the Dark Universe''), CA23130 (``Bridging high and low energies in search of quantum gravity (BridgeQG)''), CA21136 (``Addressing observational tensions in cosmology with systematics and fundamental physics (CosmoVerse)''), and CA23115 (``Relativistic Quantum Information (RQI-Action)''). 
\end{acknowledgments}

\section*{Data Availability Statement}
All computational scripts and data that support the findings of this study are available from the corresponding author upon reasonable request.

\bibliographystyle{apsrev4-2}
\bibliography{finalref}

\end{document}